\documentclass[compsoc, conference, a4paper, 10pt, times]{IEEEtran}

\usepackage{algorithmic}
\usepackage{graphicx}
\usepackage{textcomp}
\usepackage{xcolor}
\usepackage{colortbl} 
\usepackage{subcaption}
\usepackage{caption}
\usepackage{xspace}
\usepackage{booktabs}
\usepackage{amssymb}
\usepackage{multirow}
\usepackage{diagbox}
\usepackage[flushleft]{threeparttable}
\usepackage{scalerel}
\usepackage{tikz}
\usetikzlibrary{svg.path}

\definecolor{orcidlogocol}{HTML}{A6CE39}
\tikzset{
  orcidlogo/.pic={
    \fill[orcidlogocol] svg{M256,128c0,70.7-57.3,128-128,128C57.3,256,0,198.7,0,128C0,57.3,57.3,0,128,0C198.7,0,256,57.3,256,128z};
    \fill[white] svg{M86.3,186.2H70.9V79.1h15.4v48.4V186.2z}
                 svg{M108.9,79.1h41.6c39.6,0,57,28.3,57,53.6c0,27.5-21.5,53.6-56.8,53.6h-41.8V79.1z M124.3,172.4h24.5c34.9,0,42.9-26.5,42.9-39.7c0-21.5-13.7-39.7-43.7-39.7h-23.7V172.4z}
                 svg{M88.7,56.8c0,5.5-4.5,10.1-10.1,10.1c-5.6,0-10.1-4.6-10.1-10.1c0-5.6,4.5-10.1,10.1-10.1C84.2,46.7,88.7,51.3,88.7,56.8z};
  }
}

\newcommand\orcidicon[1]{\href{https://orcid.org/#1}{\mbox{\scalerel*{
\begin{tikzpicture}[yscale=-1,transform shape]
\pic{orcidlogo};
\end{tikzpicture}
}{|}}}}

\usepackage{hyperref}
\newcommand{\dns}[1]{{\small \texttt{#1}}}
\newcommand{\dnsfn}[1]{{\footnotesize \texttt{#1}}}

\newcommand{\ie}{i.e.}
\newcommand{\eg}{e.g.}
\newcommand{\etal}{et al.}

\def\BibTeX{{\rm B\kern-.05em{\sc i\kern-.025em b}\kern-.08em
    T\kern-.1667em\lower.7ex\hbox{E}\kern-.125emX}}

\renewcommand{\paragraph}[1]{\vspace{5pt}\noindent\textbf{#1:}}

\newcommand{\ltgrey}{\rowcolor[gray]{0.95}}
\definecolor{chicagomaroon}{rgb}{0.5, 0.0, 0.0}
\definecolor{midgreen}{rgb}{0.0, 0.50, 0.0}

\makeatletter
\newcommand{\linebreakand}{%
  \end{@IEEEauthorhalign}
  \hfill\mbox{}\par
  \mbox{}\hfill\begin{@IEEEauthorhalign}
}
\makeatother

\title{
  Forward Pass: On the Security Implications of\\Email Forwarding Mechanism and Policy\\
}

\begin{document}


\iffalse

\newcommand{\todo}[1]{\textsf{\textcolor{red}{[TODO: #1]}}}
\newcommand{\grant}[1]{\textsf{\textcolor{teal}{[GH: #1]}}}
\newcommand{\amirian}[1]{\textsf{\textcolor{violet}{[AM: #1]}}}
\newcommand{\gakiwate}[1]{\textcolor{violet}{\noindent[Gautam: #1]}}
\newcommand{\geoff}[1]{\textcolor{purple}{\noindent[GV: #1]}}
\newcommand{\stefan}[1]{\textcolor{green}{\noindent[SS: #1]}}
\newcommand{\mattijs}[1]{\textcolor{blue}{\noindent[MJ: #1]}}
\newcommand{\alex}[1]{\textcolor{chicagomaroon}{\noindent[AL: #1]}}
\else
\newcommand{\todo}[1]{}
\newcommand{\amirian}[1]{}
\newcommand{\grant}[1]{}
\newcommand{\gakiwate}[1]{}
\newcommand{\geoff}[1]{}
\newcommand{\stefan}[1]{}
\newcommand{\mattijs}[1]{}
\newcommand{\alex}[1]{}
\fi

\author{
\IEEEauthorblockN{Enze Liu \orcidicon{0000-0003-4288-8485}}
  \IEEEauthorblockA{
  \textit{UC San Diego}\\
  La Jolla, CA, USA \\
  e7liu@eng.ucsd.edu}
\and
\IEEEauthorblockN{Gautam Akiwate}
\IEEEauthorblockA{
\textit{Stanford University}\\
Stanford, CA, USA \\
gakiwate@cs.stanford.edu}
\and
\IEEEauthorblockN{Mattijs Jonker}
\IEEEauthorblockA{
\textit{University of Twente}\\
Enschede, Netherlands \\
m.jonker@utwente.nl}
\and
\IEEEauthorblockN{Ariana Mirian}
\IEEEauthorblockA{
\textit{UC San Diego}\\
La Jolla, CA, USA \\
amirian@cs.ucsd.edu}\\
\linebreakand
\IEEEauthorblockN{Grant Ho}
\IEEEauthorblockA{
\textit{UC San Diego}\\
La Jolla, CA, USA \\
grho@eng.ucsd.edu}
\and
\IEEEauthorblockN{Geoffrey M. Voelker}
\IEEEauthorblockA{
\textit{UC San Diego}\\
La Jolla, CA, USA \\
voelker@cs.ucsd.edu}
\and
\IEEEauthorblockN{Stefan Savage}
\IEEEauthorblockA{
\textit{UC San Diego}\\
La Jolla, CA, USA \\
savage@cs.ucsd.edu}\\
}

\maketitle

\begin{abstract}
The critical role played by email has led to a range of extension
protocols (\eg, SPF, DKIM, DMARC) designed to protect against the
spoofing of email sender domains.  These protocols are complex as is,
but are further complicated by automated email forwarding --- used by
individual users to manage multiple accounts and by mailing lists to
redistribute messages.  In this paper, we explore how such email
forwarding and its implementations can break the implicit assumptions
in widely deployed anti-spoofing protocols.  Using large-scale empirical
measurements of 20 email forwarding services (16 leading email providers and four popular mailing
list services), we identify a range of security issues rooted in
forwarding behavior and show how they can be combined to reliably
evade existing anti-spoofing controls.  We further show how these issues allow
attackers to not only deliver spoofed email messages to prominent email providers (e.g., Gmail, Microsoft Outlook, and Zoho), but also reliably spoof email on behalf of tens of thousands of
popular domains including sensitive domains used by organizations in
government (\eg, \dns{state.gov}), finance (\eg, \dns{transunion.com}), law (\eg,
\dns{perkinscoie.com}) and news (\eg, \dns{washingtonpost.com}) among others.

\end{abstract}



\section{Introduction}
Email has long been a uniquely popular medium for social engineering
attacks.\footnote{In the 2021 Verizon Data Breach Investigation
  Report, phishing is implicated in 36\% of the more than 4,000 data
  breaches investigated; and email-based attacks, including Business
  Email Compromise (BEC), completely dominate the social engineering
  attack vector~\cite{dbir2021}.}  While it is widely used for
both unsolicited business correspondence as well as person-to-person
communications, email provides no intrinsic integrity guarantees.  In
particular, the baseline SMTP protocol provides no mechanism to
establish if the purported sender of an email message (\eg, From:
\dns{Anthony.Blinken@state.gov}) is in fact genuine.

To help address this issue, starting in the early 2000's, the email
operations community introduced multiple anti-spoofing protocols,
including the Sender Policy Framework (SPF)~\cite{rfc7208}, DomainKeys
Identified Mail
(DKIM)~\cite{rfc6376} and Domain-based Message Authentication
Reporting and Conformance (DMARC)~\cite{rfc7489}, each designed to
tighten controls on which parties can successfully deliver
mail purporting to originate from particular domain names.  However,
these protocols had the disadvantage of being both post-hoc (needing
to support existing email deployments and conventions) and piecemeal
(each addressing slightly different threats in slightly different
ways).  As a result, the composition of these protocols is complex and
hard to reason about, leading to a structure that Chen \etal\ recently
demonstrated can enable a range of evasion
attacks~\cite{chen2020composition}.

In this paper, we explore the unique aspects of this problem created
as a result of \emph{email forwarding}, which is commonly used by both
individuals (\ie, to aggregate mail from multiple accounts) and
organizations (\ie, for mailing list distribution).  While clearly
useful, forwarding introduces a range of new interaction
complexities. First, forwarding involves three parties instead of two
(the sender, the forwarder, and the receiver), where the
``authenticity'' of an email message is commonly determined by the party with
the weakest security settings.  Second, the intrinsic nature of email
forwarding is to transparently send an existing message to a new
address ``on behalf'' of its original recipient --- a goal very much
at odds with the anti-spoofing function of protocols such as SPF and
DMARC.  For this reason, forwarded email messages can receive special
treatment based on various assumptions about how forwarding is used in
practice.  Finally, there is no single standard implementation of
email forwarding. Different providers make different choices and the
email ecosystem is forced to accommodate them.  Unfortunately, some
problematic implementation choices (\eg, permitting ``open
forwarding'') incur no security impact on the implementing party but
can jeopardize the security of downstream recipients.  This inversion of
incentives and capabilities creates additional challenges to
mitigating forwarding vulnerabilities.

To characterize the nature of these issues, we conduct a large-scale empirical
measurement study to infer and characterize the mail forwarding
behaviors of 16 leading email providers and four popular mailing
list services.  From these results, we identify a range of implicit
assumptions and vulnerable features in the
configuration of senders, receivers, and forwarders.  Using a
combination of these factors, we then demonstrate a series of distinct
evasion attacks that bypass existing anti-spoofing protocols and allow
the successful delivery of email with spoofed sender addresses (\eg,
From: \dns{Anthony.Blinken@state.gov}).  These attacks affect both leading
online email service providers (\eg, Gmail, Microsoft Outlook, iCloud, and
Zoho) and mailing list providers/software (\eg, Google Groups and
Gaggle).  Moreover, some of these issues have extremely broad
impact --- affecting the integrity of email sent from tens of
thousands of domains, including those representing organizations in the
US government (spanning the majority of US cabinet domains, such as
\dns{state.gov} and \dns{doe.gov}, as well as the domains of security agencies
such as \dns{odni.gov}, \dns{cisa.gov}, and \dns{secretservice.gov}), financial services
(\eg, \dns{transunion.com}, \dns{mastercard.com}, and \dns{discover.com}), news (\eg,
\dns{washingtonpost.com},
\dns{latimes.com},
\dns{apnews.com}, and \dns{afp.com}), commerce (\eg, \dns{unilever.com}, \dns{dow.com}), and law (\eg,
\dns{perkinscoie.com}).
Finally, in addition to disclosing these issues to their respective
providers, we discuss the complexities involved in identifying,
mitigating, and fixing such problems going forward.


\section{Background}
\label{sec:background}

In this section, we describe the anatomy of a simple email
transmission and the protocols used to authenticate such an email.
We also present a high-level overview of how forwarding modifies the
email delivery flow as a basis for a detailed description of
different forwarding approaches and implementations in
Section~\ref{sec:measure_forwarding_mechs_and_arc}.  Finally, we
briefly survey related work on email security, particularly
those whose insights we have built upon.

\subsection{Simple Mail Transfer Protocol}
\begin{figure}[t]
    \centerline{\includegraphics[width=\columnwidth]{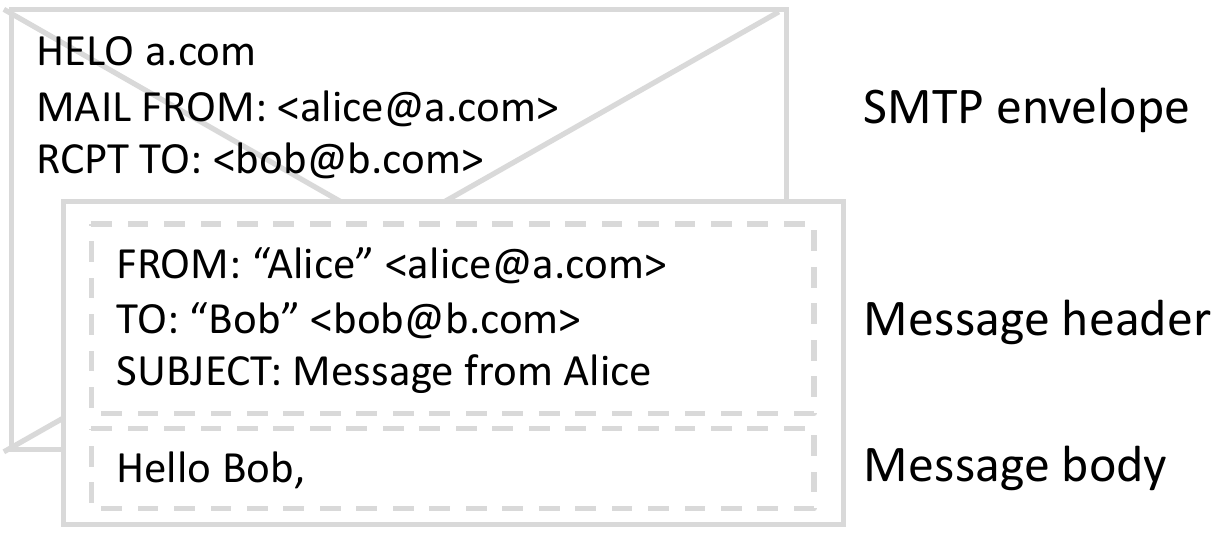}}
    \centering
    \caption{Example SMTP headers in a transmission (inspired by Figure 3 in Chen et al.~\cite{chen2020composition}).}
    \label{fig:mech_smtp_headers}
\end{figure}

The Simple Mail Transfer Protocol (SMTP) governs the addressing and
delivery of Internet email~\cite{rfc5321}.  Designed to mimic physical
mail, SMTP specifies two distinct sets of headers that declare the
sender and recipient(s) of an email message.  An outer set of headers,
the \emph{SMTP Envelope Headers} (\textsc{MAIL FROM} and \textsc{RCPT
  TO}), tell email servers how to route and deliver email. In
particular, the \textsc{RCPT TO} header identifies the message's
recipient and the \textsc{MAIL FROM} header identifies where to send
replies and bounce messages.  An inner set of headers, the
\emph{Message Headers} (\textsc{FROM} and \textsc{TO}), are contained
in the body of the SMTP message~\cite{rfc5322}.  These correspond to the
human-readable names and addresses set by email clients when the
sending user creates an email message.  These headers are strictly
intended for human user-interface purposes (\ie, for populating the
``To:'' and ``From:'' fields in email clients) and they are not used
for email routing.  Figure~\ref{fig:mech_smtp_headers} illustrates an
example message with both sets of headers.  Note that, although the
addresses in the Envelope and Message headers frequently match (as
they do in our example), they are not required to do so and there are both
benign (\eg, email forwarding) and malicious (\eg, phishing) reasons
for producing mismatched headers (\eg, where the \textsc{MAIL FROM}
address does not match the \textsc{FROM} address).

\subsection{Email Spoofing Protections}
The original SMTP design lacks authentication, which
has made email spoofing attacks both possible and common. To mitigate these attacks, the community has proposed multiple mechanisms that focus on authenticating the \emph{domain name} used by the
purported sender.\footnote{True per-sender authentication has long floundered
  due to the lack of effective mechanisms for binding user identities
  with cryptographic credentials at scale.  The best known protocol in
  this space, PGP, has been riddled with security and usability issues
  and remains, at best, a niche protocol.  In this paper, we focus
  exclusively on domain-level sender authentication.} Of these
mechanisms, we focus on SPF~\cite{rfc7208}, DKIM~\cite{rfc6376}, and DMARC~\cite{rfc7489} given their wide adoption.

\medskip
\noindent\textbf{Sender Policy Framework (SPF)} defines a list of IP addresses permitted to send email on behalf of a domain and a set of actions the recipient should take if they receive an email from an unauthorized IP address.\footnote{In addition to lists of raw IP addresses, SPF
records can also ``include'' other SPF records by reference.}
Domain owners specify this policy by publishing it in a DNS TXT record.
Upon receiving an email message, the receiver fetches the list of authorized sender IP addresses by querying the domain in the email's \textsc{MAIL FROM} header.
The recipient then verifies if the IP address of the sending server is included that list.
If the verification fails, the receiver enforces the action (e.g., marking the email as spam) specified by the \textsc{MAIL FROM} domain in their SPF policy.

%

\medskip
\noindent\textbf{DomainKeys Identified Mail (DKIM)} cryptographically
binds an email message with its sending email domain.  With DKIM, the
sender signs an email (or certain elements of an email) and attaches a digital signature via a
\textsc{DKIM-Signature} message header for future verification.
Receivers later retrieve the signer's public key (in the form of a DNS TXT record) from the domain specified in the \textsc{DKIM-Signature} header and authenticate an email's signature using that key.

Sadly, neither SPF nor DKIM verify that an email's purported sender (\ie, the \textsc{FROM} header) truly wrote and sent it~\cite{chen2020composition}.
For example, an attacker could bypass DKIM by spoofing an email's \textsc{FROM} header, but then sign and attach a DKIM signature that uses key pairs linked to their own domain (since DKIM does not compare the signature's domain against the \textsc{FROM} domain).
Attacks that exploit
this lack of \textsc{FROM} header authentication motivated the creation of DMARC.

\medskip
\noindent\textbf{Domain Message Authentication, Reporting, and
Conformance (DMARC)} combines and extends SPF and DKIM to mitigate these security issues.
Under DMARC, an email's receiver performs an ``alignment test'': checking if the
domain in the \textsc{FROM} header matches the domain name verified by
either SPF (the domain in the \textsc{MAIL FROM} header) or DKIM (the
domain in the \textsc{DKIM-Signature} header).
By default (``relaxed mode''), the alignment test only requires that the registered domains in the headers match (\ie, not the fully qualified domain name (FQDN)). However, domain owners can specify that recipients should follow the strict mode of the alignment test, which requires the \textsc{FROM} header's FQDN to exactly match the domain authenticated by SPF or DKIM.\footnote{
DMARC policy records are also stored as DNS TXT records.} 

If the email passes either SPF or DKIM authentication, and the alignment test also passes, then DMARC considers the email authenticated.
Otherwise, the receiver should implement the DMARC policy designated by the domain in the \textsc{FROM} header, selected from one of three options: \textsc{None},
\textsc{Quarantine}, or \textsc{Reject}.  A policy of \textsc{None}
specifies that an email should be delivered as normal (and thus is often used for monitoring purposes~\cite{DMARCMonitoring1,DMARCMonitoring2}), and \textsc{Reject}
specifies that the recipient mail server should drop the email without
delivering it to the user.  The \textsc{Quarantine} policy is not strictly
defined (indicating only that the message should be treated ``as
suspicious'') and allows each email provider considerable latitude in
their implementation (\eg, setting a UI indicator or placing the email
in a designated spam folder)~\cite{rfc7489}.

\begin{figure}[t]
\centerline{\includegraphics[width=\columnwidth]{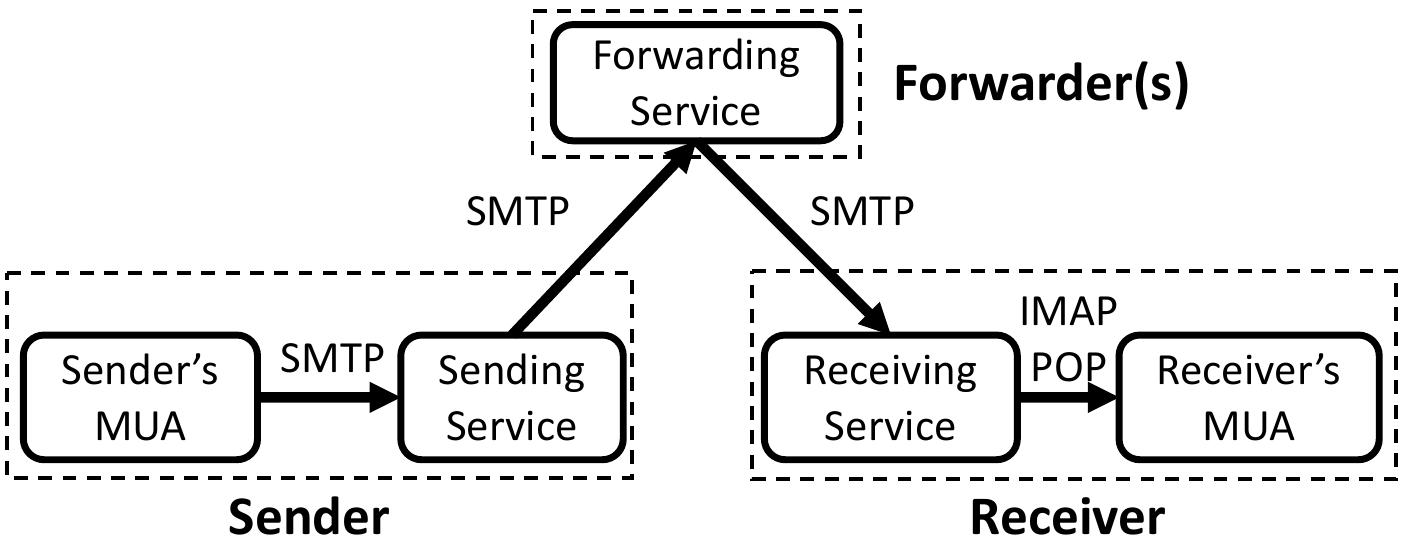}}
\caption{Email flow involving forwarding.}
\label{fig:email_forwarding_flow}
  \vspace*{-0.1in}
\end{figure}

\subsection{Email Forwarding}
\label{sec:background:fwd:overview}
Forwarding is ubiquitous in the email ecosystem and is necessitated by
the wide use of mailing lists~\cite{Electron8:online}, email filtering
services such as ProofPoint~\cite{SecureEm78:online}, and
auto-forwarding employed by individual users for account
aggregation~\cite{TheBestW9:online}, among others.  As shown in
Figure~\ref{fig:email_forwarding_flow}, forwarding alters the standard
transmission flow of an email message.  Instead of a direct
transmission from the sender to the recipient, forwarding relays an
email from the sender to an intermediate server and/or account, which
then transmits a copy of the email to the final recipient.
For simplicity we show a single forwarder in our example, but email can pass through multiple forwarders in common use cases. 

Like normal receivers in direct mail transfer,
forwarders are responsible for performing standard authentication checks on each email they receive. 
However, after authenticating a message,
a forwarder often makes \emph{changes} to the email headers and/or the
email body based on the service it provides.
The forwarder then sends the modified message to the final receiver (or next forwarder), which also performs authentication checks upon receiving the email.
Finally, when a recipient receives and opens an email, the receiver's user agent (MUA) parses and displays the message to the user.

\subsection{Related Work}

Email security has been a long-standing problem and a variety of prior research efforts have examined different aspects of it. One line of work focuses on understanding and defending against phishing attacks. This includes papers that design new tools for detecting both traditional phishing and sophisticated spearphishing attacks~\cite{abu2007comparison,bergholz2008improved, fette2007learning, garera2007framework, whittaker2010large, duman2016emailprofiler, khonji2011mitigation, zhao2016optimizing, stringhini2015ain, ho19, ho17, cidon19},
study the characteristics of real-world phishing attacks~\cite{han2016phisheye,onaolapo2016happens,thomas2014consequences,bursztein2014handcrafted}, and examine the human aspect of such attacks~\cite{lastdrager2017effective, reinheimer2020investigation,abu2007comparison,
mayer2022don,caputo2013going,
Spero20, Sheng10, Kumaraguru10}.

Another body of work investigates the security and deployment of email encryption mechanisms, such as PGP~\cite{muller2019johnny, poddebniak2018efail, schwenk2020mitigation, muller2020mailto,stransky202227}, DANE~\cite{lee2022under,lee2020longitudinal}, and STARTTLS~\cite{zakir15,Foster15,poddebniak2021tls,holz2015tls,mayer2016no}.

A third research direction analyzes the security and deployment of anti-spoofing protocols such as SPF, DKIM and DMARC, with efforts from both industry and academia. The blogposts by Ullrich~\cite{Breaking12:online} and Haddouche~\cite{Mailsplo10:online} investigated approaches for bypassing DKIM and DMARC using malformed email messages.
Other work has empirically measured the efficacy and deployment status of SPF, DKIM, and DMARC~\cite{hu_end--end_nodate, zakir15, Foster15, tatang2021evolution, deccio21, wang2022large, bennett2022spfail}, as well as qualitatively characterized the factors that drive DMARC policy decisions~\cite{hutowardsunderstanding}.

The work most related to our own includes Chen et al.'s analysis of
the security vulnerabilities introduced by protocol composition in
modern email delivery~\cite{chen2020composition}, Shen et al.'s
analysis~\cite{shen2020weak} of modern sender spoofing attacks, and Wang et al.'s~\cite{wang2022revisiting} analysis of email security under the experimental Authenticated Received Chain (ARC) protocol~\cite{rfc8617}.
Of these, Chen et al.~\cite{chen2020composition} do not consider forwarding at all and Wang et al.~\cite{wang2022revisiting} focus on ARC and only consider one specific forwarding implementation as well (REM+MOD in Section~\ref{sec:measure_forwarding_mechs_and_arc}), leaving many other vulnerable forwarding mechanisms and features unexplored.

Shen et al.'s work~\cite{shen2020weak} is the closest in that it also
examines open forwarding, but because they only consider one
forwarding mechanism (what we label as REM in
Section~\ref{sec:measure_forwarding_mechs_and_arc}), they do not
identify the significant scope of this issue.  We build on and
generalize this work to show, among other attacks, that attackers are
able to practically abuse open forwarding to spoof \emph{any} domain
that includes the forwarding domain's SPF record in their own SPF record (a
common practice when hosting email via Microsoft's Outlook service for example).



In summary, our paper builds on the insights of prior efforts, but focuses exclusively and deeply on the particular security challenges introduced by the design and features of common forwarding mechanisms, and their complex interactions with existing email protocols. Through systematic measurements and analysis, we not only show that prior work largely underestimates the risks of open forwarding,
but also reveal new attacks not discovered in prior work.




\section{Email Forwarding in Practice}
\label{sec:measure_forwarding_mechs_and_arc}

Despite the ubiquity of email forwarding, there is no single and
universally agreed-upon method for how email services should implement
forwarding, resulting in several different approaches~\cite{Emailfor34:online}.  This
heterogeneity stems in part from the difficulty of balancing
compatibility with anti-spoofing protocols and the functional goals of
many forwarding use cases: to transparently hide the intermediate
forwarder and present the illusion that the recipient receives the
email directly from its original sender.

Absent a clear standard to depend on, we have used empirical
measurements to \emph{infer} the forwarding behavior deployed by
prominent email providers and mailing list services.  For each
service, we created multiple test accounts, used them to forward email
to recipient accounts we controlled, and then analyzed the resulting
email headers to identify the forwarding mechanism employed.
(Section~\ref{sec:methodology} has a more detailed description of our
methodology.)

We constructed a comprehensive and representative set of forwarding services by building on top of prior literature.
In total, we studied 20 distinct, leading email forwarding services.
We started by collecting all email providers studied in prior literature~\cite{chen2020composition,shen2020weak,hu_end--end_nodate,wang2022revisiting}.
We considered an email provider \emph{out of scope} if it meets any of these four criteria: (a) it is no longer active (e.g., excite.com); (b) it does not accept US customers (e.g., all Chinese providers studied in prior work);  (c) it is not open to public registration (e.g., cock.li); or (d) it does not support forwarding (e.g., Protonmail).
Using these criteria, we identified 23 email providers.
Next, we excluded five email providers that prohibited bulk registration (which prevents us from running large-scale measurements), leaving us with 18 email providers.
We then identified and removed duplicate providers that are operated by the same vendor under different names, leading to a total of 14 distinct email providers.
Finally, we augmented this set of forwarding services by searching for popular email providers that supported forwarding and widely-used mailing list services (a common use case overlooked in prior literature), adding two additional email providers (Mail2World and GoDaddy) and four mailing lists.

Our selection of email forwarding services covers a diverse set of countries and real-world use cases (personal and business email), and represents services used by the general public (used by over 46\% of popular Alexa domains and government domains according to Liu et al.~\cite{liu2021s} ignoring email filtering services).
We list all email providers and mailing lists in Table~\ref{tab:forwarding_mechs_in_the_wild}.



\begin{figure}[t]
  \centering
    \includegraphics[width=\columnwidth]{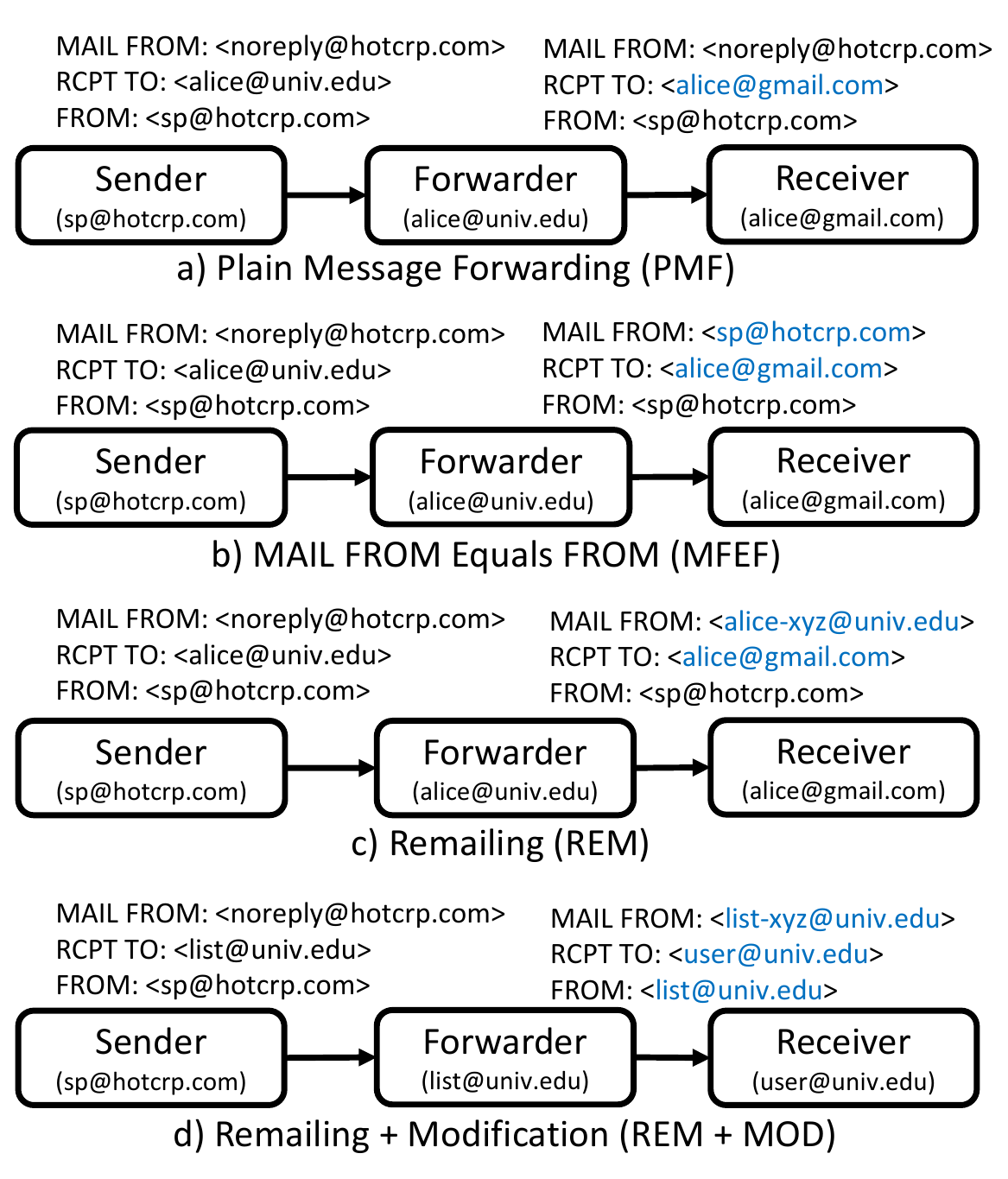}
    \caption{Four prevalent approaches to email forwarding.  Addresses
      in blue correspond to header values rewritten during the
      forwarding process.
      }
    \label{fig:forwarding_mechs_combined}
\end{figure}

Through our measurements, we confirmed the use of three common
approaches that are generally known through public documentation, and
identified a fourth uncommon implementation used by
Microsoft Outlook (hence referred to as Outlook) and Freemail.hu (hence referred to as Freemail).
As summarized in
Figure~\ref{fig:forwarding_mechs_combined}, in each approach the
forwarder modifies the sender and recipient fields in the SMTP
Envelope and Message headers before relaying the email to its
recipient.


We now describe each of these approaches in detail using two running
examples of common email forwarding use cases.  In the first case,
Alice has configured her university account (\dns{alice@univ.edu}) to
forward to her primary personal account (\dns{alice@gmail.com}).  When
her university account receives email (\eg, from
\dns{sp@hotcrp.com}), forwarding retransmits it to
\dns{alice@gmail.com} in a way that makes it seem like the email comes
directly from the sender (\dns{sp@hotcrp.com}), rather than from her
university account.  In the second case, Bob sends an email to a
mailing list (\dns{list@univ.edu}), which redistributes (forwards) the
email to the list's members (\eg, \dns{user@univ.edu}).

\paragraph{Plain Message-Forwarding (PMF)}
Initially designed for
the purpose of ``source-routing''~\cite{Emailfor45:online}, PMF was
one of the first forwarding mechanisms in wide use.
Forwarders that use PMF only change
the \textsc{RCPT TO} header from the forwarder's email account (\dns{alice@univ.edu}) to the final recipient's address (\dns{alice@gmail.com}), and leave all other fields
untouched,
as illustrated in Figure~\ref{fig:forwarding_mechs_combined}a.
This approach achieves the goal of transparent forwarding.
Changing the \textsc{RCPT TO} header will tell mail servers to send the email to the new address's account, and leaving the \textsc{FROM} header intact will
cause the recipient's email client to display the initial sender (\dns{sp@hotcrp.com}),
rather than presenting \dns{alice@univ.edu} as the sender.

\paragraph{MAIL FROM Equals FROM (MFEF)}
Similar to PMF, MFEF (Figure~\ref{fig:forwarding_mechs_combined}b) aims to achieve transparent forwarding by preserving the original sender's identity in the \textsc{FROM} header.
Unlike the other forwarding approaches described in this section, MFEF is a custom forwarding implementation that appears to be used only by Outlook and Freemail.
A MFEF forwarder not only rewrites the \textsc{RCPT TO} header to the final recipient (\dns{alice@gmail.com}), but it \emph{also} sets the \textsc{MAIL FROM}
header to be the same as the \textsc{FROM} header (from \dns{noreply@hotcrp.com} to \dns{sp@hotcrp.com}).

Email forwarded using PMF and MFEF often break SPF validation because the \textsc{MAIL FROM} domain typically does not list the forwarding server's IP address in its SPF allowlist;
in our example, \dns{hotcrp.com} does not list the email servers for \dns{univ.edu} in its SPF allowlist.
This incompatibility has hindered the adoption of SPF and DMARC~\cite{hutowardsunderstanding},
leading to provider-specific defenses and new anti-spoofing protocols that we describe in Section~\ref{sec:assumptions}.




\paragraph{Remailing (REM)}
Unlike PMF and MFEF, remailing (aka redistribution)
works well with SPF because this approach alters the headers in a way that resembles the action of the forwarder submitting a new message~\cite{SenderRe69:online}.
As shown in Figure~\ref{fig:forwarding_mechs_combined}c, the REM
forwarder (\dns{univ.edu}'s mail server) first changes the
\textsc{RCPT TO} header to specify the final recipient (\dns{alice@gmail.com}).
Additionally, the forwarder rewrites the \textsc{MAIL FROM} header so that it
corresponds to an address in the forwarder's own domain (\eg,
\dns{alice-xyz@univ.edu}).\footnote{The Sender Rewriting Scheme
(RFC~5231~\cite{rfc5231})provides a generic framework for how forwarders should
rewrite the \textsc{MAIL FROM} header.  However, email providers do not
strictly follow this scheme and the exact email address after rewriting varies
by implementation.}

However, even though REM interoperates with SPF, it can still fail DMARC
authentication.  Absent a valid DKIM header, email messages forwarded
via REM will fail DMARC's alignment test because the \textsc{FROM}
domain will not match the SPF-verified \textsc{MAIL FROM}
domain.\footnote{Many domains still do not implement DKIM for outbound email, and even
  those that do can have their user's DKIM signatures invalidated by mailing
  list software that adds content to a user's post~\cite{rfc6783}.}
This incompatibility has led to the common adoption of weaker DMARC policies,
such as \textsc{None} and \textsc{Quarantine} instead of
\textsc{Reject}~\cite{hutowardsunderstanding}.


\paragraph{Remailing with Modification (REM + MOD)}
The final forwarding approach, Remailing with Modification (REM + MOD)~\cite{rfc6783},
resolves these compatibility issues by sacrificing the goal of transparent forwarding.
Email forwarded using REM + MOD will pass both SPF and DMARC.
However, email forwarded with this approach will display the \textit{forwarder} as the email's sender to the final recipient (hiding the identity of the original sender).
Because of this functional change, most major email platforms do not adopt this approach, and it is used primarily by mailing list services such as Gaggle.

As shown in Figure~\ref{fig:forwarding_mechs_combined}d, with REM + MOD the forwarder modifies the headers just like it would during REM forwarding: changing the \textsc{RCPT TO} header to the final recipient (\dns{user@univ.edu}) and the \textsc{MAIL FROM} header to an address in the forwarder's domain.
Additionally, the forwarder rewrites the \textsc{FROM} header to match its account or an email address within its domain (e.g., \dns{list@univ.edu}).

Although this forwarding approach produces email messages compatible with DMARC,
we found that it also introduces a new set of security concerns and spoofing attacks (\S~\ref{subsec:attack_none_mailing_list}).
At a high-level, because REM + MOD rewrites a forwarded email's headers to always pass SPF and DMARC checks, it enables an attacker to launder a spoofed email through a vulnerable forwarder such that it appears like a legitimate email message to the recipient.

\begin{table}[t]
  \centering
  \begin{tabular}{ll|ll}
  \toprule
\multicolumn{2}{l}{\textbf{Email} \hfill \hspace*{0.22in}\textbf{Forwarding}} & \textbf{Mailing} & \textbf{Forwarding} \\
\multicolumn{2}{l}{\textbf{Provider} \hfill \hspace*{0.05in}\textbf{Mechanism}} & \textbf{List Service} & \textbf{Mechanism} \\
  \midrule
  Fastmail        & PMF  & Gaggle & REM+MOD \\
  Freemail.hu     & MFEF & Google Groups & REM\\
  GMX/Mail.com    & REM  & Mailman & REM  \\
  Gmail           & REM  & Listserv & REM \\
  GoDaddy         & REM  & & \\
  Hushmail        & PMF  & & \\
  iCloud          & PMF  & & \\
  Inbox.lv        & REM  & & \\
  Mail.ru         & PMF  & & \\
  Mail2World      & PMF  & & \\
  Onet.pl/Op.pl   & REM  & & \\
  Outlook/Hotmail/O365 & MFEF & & \\
  Pobox           & REM  & & \\
  Runbox          & PMF  & & \\
  Yahoo           & PMF  & & \\
  Zoho            & REM  & & \\
  \bottomrule
  \end{tabular}
  \caption{The providers and mailing list services we tested and the forwarding mechanisms they use. For providers that are operated by the same vendor under different names (e.g., GMX and Mail.com), we merge them into one row. O365 stands for Office 365.
    \label{tab:forwarding_mechs_in_the_wild}}
  \end{table}




Table~\ref{tab:forwarding_mechs_in_the_wild} summarizes the default forwarding approach used by each of the email providers and mailing lists in our study.\footnote{A provider might forward differently when forwarding between internal accounts, and a mailing list might switch to a different forwarding mechanism to avoid issues caused by forwarding email messages from domains with stricter DMARC policies~\cite{spamreso59:online}. We do not consider these two cases.}
%
%
The most common forwarding approach is remailing forwarding (REM),
used by seven email providers (GMX, Gmail, GoDaddy, Inbox, Onet, Pobox, and Zoho) and
three mailing lists (Google Groups, Listserv, and Mailman). Seven email providers, Fastmail, Hushmail, iCloud, Mail.ru, Mail2World, Runbox, and Yahoo, use
plain-message forwarding (PMF).
Outlook and Freemail use their own custom
forwarding mechanism (MFEF) and, as described, Gaggle uses remailing with modification (REM+MOD)
forwarding.

\begin{table*}[t]
    \centering
    \begin{tabular}{llll}
        \toprule
&     \textbf{Security Assumption or Feature} & \textbf{Implementation Aspect}  & \textbf{Prevalence}    \\
    \midrule
    \S~\ref{subsubsec:dmarc_none} &    Domain will use actionable DMARC policies   & DMARC None & Two-thirds of Alexa Top 1M\\
    \S~\ref{subsubsec:spf_incorporation} &  Each domain uses its own infrastructure  & Shared SPF record    & All providers \\
    \S~\ref{subsubsec:quarantine_instead_of_reject}    & Quarantining is sufficient    & Quarantine instead of reject    & Outlook, Fastmail, GMX, Inbox.lv, Pobox \\
    \S~\ref{subsubsec:whitelist}    & Per-user DMARC overrides are fate-sharing    & Domain whitelisting    & All providers  \\ [0.15in]

    \S~\ref{subsubsec:open_forwarding}    & Users only forward to accounts they control   & Open forwarding    & Ten providers including Outlook and Fastmail\\
    \S~\ref{subsubsec:relaxed_validation}    & Forwarded email from large providers benign  & Relaxed validation & Gmail, Outlook, Mail.ru   \\
    \S~\ref{subsubsec:unsolicited_dkim}    & Adding DKIM signature increases deliverability & Unsolicited DKIM signatures & iCloud, Runbox,  Hushmail\\

    \bottomrule
    \end{tabular}
    \captionof{table}{Summary of vulnerable security assumptions and forwarding features, the aspect of their implementation that leads to the vulnerability, and the prevalence of the vulnerability.
    }
    \label{tab:assumptions_and_prevalence}
    \end{table*}

\section{Assumptions and Vulnerable Features}
\label{sec:assumptions}

In this section, we describe a range of email design and
implementation weaknesses that lead to forwarding vulnerabilities.
We start by exploring four assumptions made by anti-spoofing mechanisms
that email forwarding can bypass and violate.  We then examine
three vulnerable forwarding features in the major forwarding
approaches.

In each of these cases, we use active
measurements --- either of mail services themselves or the DMARC policies
as stored in DNS --- to document the prevalence of each issue
among prominent domains and
providers (summarized in Table~\ref{tab:assumptions_and_prevalence}).
In the remainder of this section, we discuss the measurement
methodology used to investigate and identify these issues and
describe each vulnerability in turn.  In the next section,
we then show how these vulnerabilities can
be combined to create complete and effective spoofing attacks
involving a broad array of popular and sensitive domains.




\subsection{Methodology}
\label{sec:methodology}


For our experiments we created test \textit{forwarding} accounts on all 20 forwarding services,
test \textit{recipient} accounts on all 16 major
email providers, and mail servers for domains we control as
the \textit{sending} accounts.
%
%
For Google Groups and Gaggle, we created mailing lists under our
university's existing service and at \dns{gaggle.email}, respectively.
The other two mailing list services (Listserv and Mailman) rely upon a
third-party backend mail server; we used
Postfix~\cite{ThePostf34:online} as the backend with DMARC
enforced. We then created mailing lists under new domains we acquired
for testing (\eg, \dns{list@listserv.ourdomain.com}).


For each combination of forwarding and recipient accounts, we sent
email using three different control domains in the FROM headers, each
with the same SPF configuration but with distinct DMARC policies:
\textsc{None}, \textsc{Quarantine}, and \textsc{Reject}.
Some services (\eg, Gmail and Outlook) will mark email messages sent
from new domains as spam until there is sufficient user interaction
with those messages.  To avoid this startup effect, we ``warmed up''
our domains using a series of legitimate exchanges.  In particular,
from each domain, we sent legitimate (\ie, unspoofed) email that
passed SPF, DKIM and DMARC to our accounts at each provider.  Any
message that was delivered to the spam folder we manually marked as
``not spam''.  After this warm up period, we validated that legitimate
(\ie, unspoofed) email from our domains was properly delivered to
account inboxes in all cases.

Having primed our accounts, we assessed the prevalence of each vulnerability by
sending legitimate and spoofed email messages to all
pairwise combinations of our forwarding and recipient accounts.\footnote{Our code for automatically sending these messages is available upon request.}  We
analyzed the headers and outcomes of these attempts, and recorded
which parties exhibited vulnerable behavior.  In particular, we
configured all forwarders to forward email messages to all receivers
and recorded whether each message was delivered to the inbox, spam
folder, or rejected without delivery by each receiver.  We also noted
whether any UI warning was shown in the native web-based MUA.

\subsection{Email Security Assumptions}
\label{subsec:assumptions}
Anti-spoofing mechanisms define a set of validation procedures which
both explicitly and implicitly rely on assumptions about the behavior
of domain holders, email providers and users.  Here we identify four
such assumptions that are crucial to these defenses in the direct
single-hop delivery context, but do not necessarily hold in the
presence of email forwarding.



\subsubsection{Domains use actionable DMARC policies}
\label{subsubsec:dmarc_none}
DMARC enables recipients to authenticate whether an email truly
originates from its purported sending domain.  However, when a
recipient encounters a spoofed or illegitimate email that fails
authentication, DMARC relies on the true domain owner to specify a
policy for how to treat such email.  This design assumes that domain
owners will use DMARC policies that result in protective
actions, such as \textsc{Quarantine} or \textsc{Reject}.  When a
domain owner chooses a weaker policy, mail providers deliver the
illegitimate email to a user's inbox even if the DMARC authentication
fails, in accordance with email standards (RFC 7489~\cite{rfc7489}).
Unfortunately, prior work has shown that a large number of domains use
weak DMARC policies of
\textsc{None}~\cite{hu_end--end_nodate,tatang2021evolution,hutowardsunderstanding,
  secplaintxt, maroofi2020defensive, adoptionofschemes}, with roughly
two-thirds of the Alexa Top 1M domains employing such a policy (as of
May 2020).  While poor security hygiene accounts for some of this
outcome, many domains choose a weak DMARC enforcement policy for
deliverability concerns due to incompatibility with forwarding~\cite{hutowardsunderstanding}.

%

Cognizant of this reality, several major email providers have decided to take two types of security actions against email that fails DMARC authentication, regardless of the domain owner's specified policy.
First, as noted in prior work~\cite{hu_end--end_nodate} and confirmed in our own experiments, Outlook quarantines email if it fails DMARC authentication, even when the email's \textsc{FROM} domain has a weak DMARC policy of \textsc{None}.
Second, although Gmail, Onet, and Zoho deliver email that fails DMARC authentication to user inboxes, they will display a UI warning to users who read such messages.

These defenses provide protection against attackers who directly send spoofed email to their victims.  However, as we will show,
email forwarding introduces new complexity that enables attackers to bypass these ad hoc defenses, and thus leverage weak DMARC policies to successfully spoofed email from prominent domains.

%
%

\subsubsection{Each domain uses its own infrastructure}
\label{subsubsec:spf_incorporation}
The SPF protocol predates the emergence of large third-party email providers.
As a result, SPF implicitly assumes that each organization (domain) maintains its own mailing infrastructure: that the set of authentic server IP addresses specified by a domain's SPF record is not also used by other domains or external users to send email.
Unfortunately, as documented by Liu \etal~\cite{liu2021s} and Holzbauer \etal~\cite{holzbauer2022not}, this assumption is invalid today as many organizations outsource their email infrastructure to the \emph{same} third-party providers such as Outlook and Gmail.  Hence, all of these domains have delegated the right to send on their behalf to the same third-party --- trusting that they will ensure isolation in spite of this blanket authorization.


Concretely, our measurements show that all 16 email providers in our study appear to configure their email infrastructure in this shared fashion.
Additionally, at least for email messages forwarded in our experiments, all providers but one (Fastmail) use the same set of servers to send both direct email and forwarded email.

Since SPF no longer provides isolation in this model, the email
providers in our study effectively \emph{simulate} it by preventing users
from setting arbitrary values in their \textsc{FROM} header.  Thus,
even though each mail provider is empowered to send any email on
behalf of all their mail customers, they prevent customers from taking
advantage of this situation by \emph{internally} restricting the \textsc{FROM} headers of outbound email messages coming from a customer's domain.
While this defense is effective in the absence of forwarding, we will
show how open forwarding mechanisms bypass this filtering (by generating spoofed \textsc{FROM} headers from an \emph{external} server controlled by the adversary),
exposing the latent conflict between SPF's design and modern mail service
use---ultimately allowing unrestricted email spoofing.


\subsubsection{Quarantining is sufficient}
\label{subsubsec:quarantine_instead_of_reject}
\begin{figure}[t]
    \centering
{
    \setlength{\fboxsep}{0pt}
    \setlength{\fboxrule}{0.5pt}
    \fbox{\includegraphics[clip,width=\columnwidth]{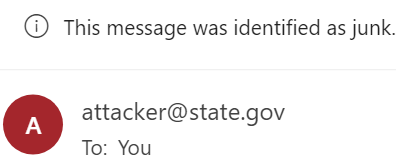}}
}
    \vspace*{-0.2in}
    \caption{Example message with a FROM header spoofing a domain with DMARC policy \textsc{Reject}.  Outlook delivers it to the spam folder instead of rejecting it.
    }
    \label{fig:example_ms_not_rejecting}
    \end{figure}

RFC 7489~\cite{rfc5231} suggests that if an email message falls under the scope of a DMARC reject policy, then the receiving server should reject and drop it entirely. However, some providers deviate from this advice by marking it as spam and delivering it to a spam folder, assuming that quarantining a malicious email neutralizes its threat.
Our experiments found that five email providers (Outlook, Fastmail, GMX, Inbox.lv and Pobox) adopt this approach.
Figure~\ref{fig:example_ms_not_rejecting} displays an email message from our tests that shows this behavior: it fails DMARC validation, comes from a domain (\dns{state.gov}) that has a DMARC policy of \textsc{Reject}, but is nonetheless delivered
as ``spam''.

%

Because these providers quarantine the spoofed email as spam, this
design does not appear particularly dangerous.\footnote{Some in the
  mail security industry criticize this weakening of DMARC
  rules and document attacks that ``rescue'' such email from the spam folder via social engineering~\cite{Microsof7:online,Spearphi83:online}.} However, as we
will show, in combination with email forwarding and another
vulnerable feature (per-user domain whitelists in Section~\ref{subsubsec:whitelist}), attackers can override this
protection and exploit the quarantine-over-reject implementation to spoof
email from thousands of popular domains despite their strict
DMARC \textsc{Reject} policy.



%

\subsubsection{Per-user DMARC overrides are fate-sharing}
\label{subsubsec:whitelist}
Many email providers allow users to override DMARC decisions: users can whitelist domains, and as a result they will still deliver or forward email even if it fails DMARC.
Providers offer this flexibility because it can help mitigate
errors and improve mail deliverability for the users who need it.
However, this feature implicitly assumes that this approach is
fate-sharing --- that when a user overrides DMARC decisions, the risks
of that choice are localized to the individual user account.  While
true in the single-hop context, forwarding again undermines this
assumption.  If adversaries can override DMARC decisions on a
forwarding account they control, they can use that capability to
launder spoofed mail and successfully deliver it downstream.


Based on our measurements, all mail providers support this functionality in some form.
Of particular note, four of the five providers 
mentioned in Section~\ref{subsubsec:quarantine_instead_of_reject}
(Fastmail, GMX, Inbox.lv, and Pobox)
allow users to override any DMARC decision for any domains.
The fifth (Outlook) allows users to override DMARC decisions for most domains, except for a small set of frequently-spoofed domains that have DMARC policy reject (e.g., \dns{aa.com}) where Outlook appears to apply additional, special protection mechanisms.

For Gmail, Hushmail, iCloud, Mail.ru, Onet, and Zoho, users can override DMARC decisions for domains with DMARC policy \textsc{None} or \textsc{Quarantine}, but not \textsc{Reject}.  Finally, for Yahoo, we can only override DMARC decisions for domains with a policy of \textsc{None}.






\subsection{Vulnerable Forwarding Features}
\label{subsec:fwding_vuln}
In the absence of forwarding, the assumptions described above are
largely benign and allow the effective blocking of many spoofing
attacks.  However, when combined with three vulnerable forwarding
features, open forwarding, relaxed validation, and unsolicited DKIM signatures, the weaknesses in
these assumptions permit several opportunities for bypassing DMARC's
protections.



\subsubsection{Open Forwarding}
\label{subsubsec:open_forwarding}
Many email service providers support a mechanism to automatically
forward a user's messages to another account (\eg, to aggregate mail
sent to multiple addresses into a single inbox).  Because of the
prevalence of these common, benign forwarding use cases, many
platforms follow a design that we call \emph{open forwarding} (also
referred to as ``unauthorized forwarding'' in previous work~\cite{shen2020weak}).  Services with open forwarding allow users
to configure their account to forward messages to any destination
email address, \textit{without} any verification from the destination
address.  Open forwarding implicitly assumes users will only forward
email to accounts that they control or have a benign relationship with
(an assumption that fails when an adversary creates or controls an account
entirely for the purpose of malicious forwarding).




Our measurements show that open forwarding is still prevalent among providers.
Specifically, ten email providers (Outlook, Fastmail, iCloud, Freemail, GoDaddy, Hushmail,
Mail2World, Onet, Pobox, and Runbox) allow open
forwarding.\footnote{Mail2World and Pobox do notify the destination account
via email about the forwarding setup.}  Moreover, as we demonstrate in three
attacks described in
Sections~\ref{subsec:attack_open_forwarding}--\ref{subsec:attack_zoho_arc},
when combined with other vulnerabilities, adversaries can exploit open
forwarding to attack not only users on those providers that employ
this design, but also a broad array of users on other platforms that
disallow open forwarding.

\begin{table*}[t]
  \centering
  \begin{threeparttable}

  \begin{tabular}{llll}
    \toprule
    & \textbf{Send email spoofing} & \textbf{Forward via} & \textbf{Deliver to} \\
    \midrule
   \multirow{2}{*}{\S~\ref{subsec:attack_open_forwarding}} & Domains with the forwarding domain's SPF information & \multirow{2}{*}{Six providers including Outlook and iCloud} & \multirow{2}{*}{Any recipient} \\
    & in their SPF records & & \\[1pt]
  \ltgrey
  & Arbitrary domains with DMARC policy None or Quarantine & Outlook & Gmail \\[1pt]
  \ltgrey
    \multirow{-2}{*}{\S~\ref{subsec:attack_relaxed_forwarding_validation}} & Arbitrary domains with DMARC policy None
    & Multiple providers (e.g., Fastmail)& Outlook \\[1pt]
    \S~\ref{subsec:attack_zoho_arc}\tnote{*} & Arbitrary domains & Fastmail & Zoho \\[1pt]
  \ltgrey
    & Domains hosting the mailing list and DMARC policy None & Google Groups, Listserv, Mailman & Any recipient \\
  \ltgrey
    \multirow{-2}{*}{\S~\ref{subsec:attack_none_mailing_list}} & Arbitrary domains & Gaggle & Any recipient \\
    \bottomrule
  \end{tabular}

  \begin{tablenotes}
  \item[*] We build on the ARC vulnerability
    identified by Shen et al.~\cite{shen2020weak}, to demonstrate an
    attack that is practical.
  \end{tablenotes}

  \end{threeparttable}
  \caption{Summary of email forwarding attacks (\S~\ref{sec:attacks}).
    \label{tab:summary_attacks}} 

  \end{table*}

\subsubsection{Relaxed Validation}
\label{subsubsec:relaxed_validation}
Since forwarded email can break SPF and DMARC at times, providers may employ relaxed validation for email forwarded by large email providers, assuming that these large providers will
prevent spoofed email messages from being forwarded.\footnote{Shen et al.~\cite{shen2020weak} also make this observation, but do not document the concrete steps necessary to exploit this vulnerability or demonstrate its practical exploitation.}


We infer that three providers, Gmail, Outlook and Mail.ru, apply some form of relaxed validation. Gmail employs two versions of relaxed validation for forwarded email messages that both (1) fail SPF and DMARC checks and (2) are from domains with a DMARC policy of \textsc{None} or \textsc{Quarantine}.
First, for email messages forwarded via Gmail or Outlook, Gmail delivers them regardless.
Second, for messages forwarded via
the other providers in our experiments,
Gmail delivers the email if it meets specific conditions (more details in Appendix~\ref{sec:append_change_behavior_details}).

Similarly, our experiments found that Outlook applies relaxed validation for email messages from domains with a DMARC policy of \textsc{None}
(as discussed in Section~\ref{subsubsec:dmarc_none}, Outlook usually overrides the policy of \textsc{None} and quarantines messages that fail DMARC).
Specifically, Outlook accepts email messages forwarded via nine major providers (e.g., Gmail and Fastmail),
despite failing SPF and DMARC checks.
Finally, Mail.ru accepts email messages forwarded via Gmail that fail DMARC from domains with a DMARC policy of \textsc{None} or \textsc{Quarantine}.


These relaxed validation policies aim to balance the incompatibility of forwarding approaches with anti-spoofing protocols by implicitly trusting high-profile email services.
Unfortunately, the complexity introduced by forwarding and its interactions with the diverse set of assumptions we highlight enable attackers to abuse these trust relationships.  This is particularly true because all of these providers offer individual consumer accounts.
For example, in Section~\ref{subsec:attack_relaxed_forwarding_validation} we show that an adversary can deliver spoofed email messages from domains that have a DMARC policy of \textsc{None} or \textsc{Quarantine} to any Gmail user without triggering a warning.

\subsubsection{Unsolicited DKIM Signatures for Hosted Domains}
\label{subsubsec:unsolicited_dkim}
RFC 6376~\cite{rfc6376} and RFC 6377~\cite{rfc6377} both recommend that
forwarding services apply their own DKIM signatures for forwarded email
messages, especially for cases where they modify the message. 
Shen et al.~\cite{shen2020weak} showed that this configuration can be exploited
by a malicious actor via an attack that they called the DKIM Signature Fraud
Attack.  Specifically, they showed that an adversary can acquire valid DKIM
signatures for spoofed email messages if the forwarder naively signs every
forwarded email. Such spoofed email messages can successfully pass subsequent
DMARC checks if their spoofed sender's domain is the same as the domain used by
the forwarding service to sign DKIM signatures. Shen et al.~\cite{shen2020weak}
found three providers that had this vulnerable feature: Yahoo, Office365 and
Alibaba Cloud.

Through our experiments, we identified that three providers' (iCloud, Hushmail, and Runbox) forwarding implementation contained a variant of this vulnerable feature, which would allow an adversary to mount attacks similar to the DKIM Signature Fraud Attack.
Taking iCloud as an example, we find that iCloud adds unsolicited and valid DKIM signatures to spoofed email messages addressed from domains hosted by them. Additionally,
iCloud signs the DKIM signature using the same domain as the purported sender's domain in the spoofed email. For instance, iCloud will add a valid DKIM signature signed by the domain \texttt{peterborgapps.com} (a domain hosted by iCloud) to spoofed email messages purporting to be from \texttt{peterborgapps.com}, allowing the spoofed email messages to pass subsequent DMARC checks.
We surmise that providers can add valid DKIM signatures on behalf of hosted domains because they manage DKIM keys for these domains~\cite{Setupane66:online, HushDKIM}.

\section{Attacks}
\label{sec:attacks}
\begin{figure}[t]
  \centering
{
    \setlength{\fboxsep}{0pt}
    \setlength{\fboxrule}{0.5pt}
    \fbox{\includegraphics[trim=2 2 2 47,clip,width=\columnwidth]{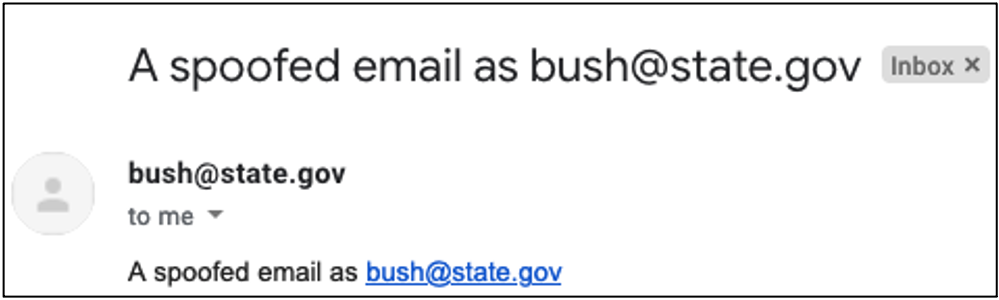}}
}
  \caption{Example of a successful attack. A spoofed email purporting to be \dns{bush@state.gov} is delivered to a Gmail user's inbox with no warning indicators.
}
\label{fig:open_forwarding_attack_screenshot}
\end{figure}

In this section, we demonstrate how an adversary can combine and
exploit the issues
described in Section~\ref{sec:assumptions} to create attacks that reliably
bypass existing anti-spoofing protections.  In particular, we consider an
attack successful if a spoofed email message is delivered to a
victim's inbox (\ie,  not the spam folder), and yet does not produce a
warning to the user.
Figure~\ref{fig:open_forwarding_attack_screenshot} shows
an example of a successful attack, where a spoofed
email purporting to be from \dns{bush@state.gov} is delivered to a
Gmail user's inbox with no warning indication.

We describe four distinct classes of attacks, summarized in
Table~\ref{tab:summary_attacks}, each of which we have validated
empirically using accounts created at the affected providers.  Some of
these attacks are quite broad --- allowing an attacker to spoof email
to any email recipient purporting to be from tens of thousands of
popular and sensitive domains --- while others are more circumscribed
in their impact.
For each of the attacks described below, we refer to the domain an
attacker specifies in their \textsc{FROM} header as the
\textit{spoofed domain}.  We use the terms \textit{spoofed address} to
refer to the full email address appearing in the \textsc{FROM} header
and \textit{forwarding domain} to refer to the domain of the
forwarder.

\begin{figure*}[t]
  \centerline{\includegraphics[width=\textwidth]{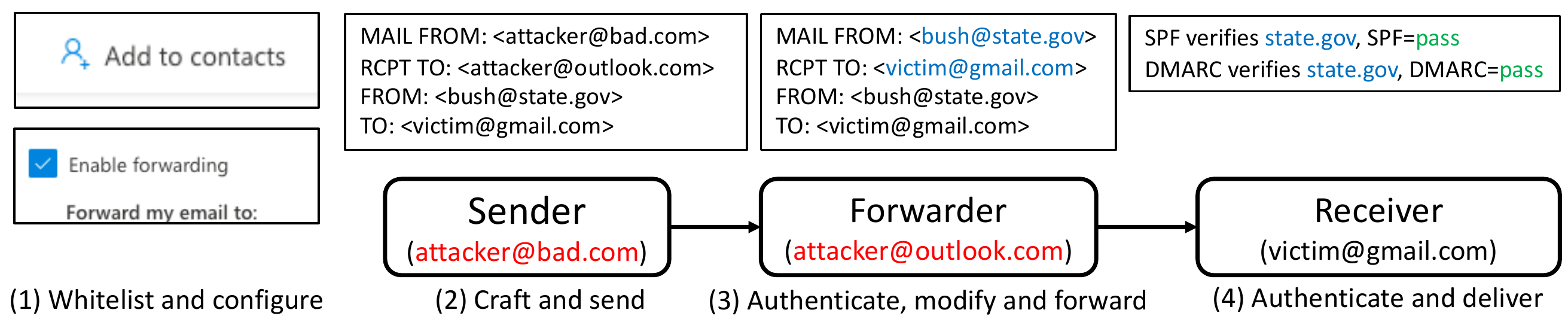}}
  \centering
  \caption{Example of an SPF Incorporation Attack (\S~\ref{subsec:attack_open_forwarding}) exploiting Outlook's open forwarding to spoof email from domains incorporating Outlook's SPF records (e.g., \dns{state.gov}) to arbitrary recipients.}
  \label{fig:open_forwarding_attack_mechanism}
\end{figure*}

\paragraph{Threat Models}
For the first three attacks, we assume an adversary controls the
sender and forwarding accounts: they possess a server capable of
sending spoofed email messages (sender) and a personal account with a
specific third-party provider that allows \emph{open forwarding}
(forwarder). For the attack described in
Section~\ref{subsec:attack_none_mailing_list}, we make three
assumptions: (a) that adversaries control a malicious server that can
send spoofed email messages and try to spoof email from a domain that
hosts a mailing list with REM forwarding (e.g., Google Groups,
Listserv and Mailman as described in
Section~\ref{sec:measure_forwarding_mechs_and_arc}), (b) that the spoofed domain
has a DMARC policy of \textsc{None} (all too common); and (c) the
sending email address the attacker wishes to impersonate has
permission to send to the mailing list.

\begin{figure*}[t]
  \centerline{\includegraphics[width=\textwidth]{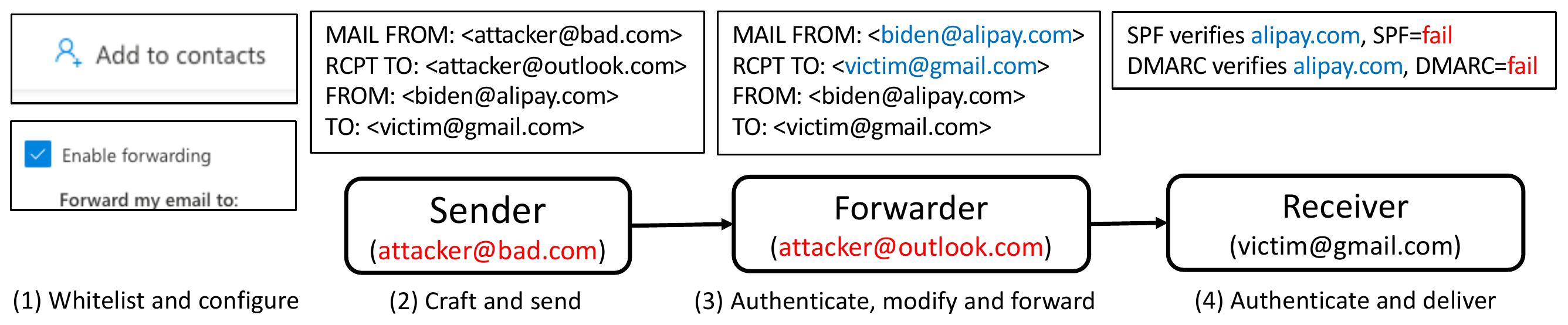}}
  \centering
  \caption{Example of a spoofed email attack exploiting open forwarding and relaxed validation for forwarded email from well-known providers (\S~\ref{subsec:attack_relaxed_forwarding_validation}).
    Note that the spoofed domain, \dns{alipay.com}, has a DMARC policy of Quarantine and thus
    should not be delivered.}
  \label{fig:mech_gmail_via_outlook}
\end{figure*}

\subsection{Exploiting SPF Incorporation}
\label{subsec:attack_open_forwarding}

The first attack we describe exploits five discrete issues:
three security assumptions (\S~\ref{subsubsec:spf_incorporation},~\ref{subsubsec:quarantine_instead_of_reject},~\ref{subsubsec:whitelist}),
the vulnerable \emph{open forwarding} feature that many providers offer (\S~\ref{subsubsec:open_forwarding}),
and the header rewriting performed as part of the PMF and MFEF forwarding approaches (\S~\ref{sec:measure_forwarding_mechs_and_arc}).
Crucially,
the rise of large third-party email providers violates SPF's assumption that the set of authorized server IP addresses specified by each domain cannot be used by other domains or external users to send email.
For example, the owners of domain \dns{state.gov} use Outlook as their email provider.  Thus, email messages sent by \dns{state.gov}'s employees will originate from Outlook's mail servers.
To ensure reliable delivery, such domains routinely add the server IP addresses of their email provider to their own SPF records.
Although intuitive, this configuration creates an overly broad trust assumption: by adding the provider IP addresses to their SPF record, such domains (e.g., \dns{state.gov}) implicitly grant permission for any account hosted by their provider, whether individual or corporate, to send email messages that purportedly come from their domain.
This threat is only prevented because large providers like Outlook do not allow users to arbitrarily set or forge their email's FROM header.

However, we observe that by combining header rewriting from PMF and MFEF and the use of \emph{open forwarding}, attackers can overcome this defense and exploit SPF's violated assumption.
Specifically, this attack allows an adversary to spoof email from domains that incorporate a third-party provider's SPF information in their own SPF record to any recipient, regardless of the domain's DMARC policy.

\paragraph{Scope}
This attack works for domains that include the SPF record of any of six
large email providers (Outlook, iCloud, Freemail, Hushmail, Mail2World and Runbox) in their own SPF records.  Notably, given
Outlook's importance as a third-party provider~\cite{liu2021s}, this
attack allows an attacker to spoof email on behalf of tens of
thousands of popular domains.

Indeed, over 12\% of the Alexa 100K
most popular domains are vulnerable as a result (and almost 8\% of the
top 1M domains).  A cursory examination of this list identified a
range of potentially sensitive domains such as those hosting large
news reporting organizations (\eg, \dns{washingtonpost.com},
\dns{latimes.com},
and \dns{apnews.com}),
financial services (\eg,
\dns{mastercard.com}, \dns{transunion.com},
and \dns{docusign.com}),
domain registrars (\eg, \dns{godaddy.com}),
certificate authorities (\eg,
\dns{sectigo.com} and \dns{digicert.com}) and large law firms (\eg,
\dns{perkinscoie.com}).  In addition, 32\% of US \dns{.gov} domains are
vulnerable (including 22\% of the domains used by Federal
agencies).  At the Federal level this includes the majority of US
cabinet organizations (\eg, \dns{state.gov}, \dns{dhs.gov} and
\dns{doe.gov}), a range of security sensitive agencies (\eg,
\dns{odni.gov}, \dns{cisa.gov} and \dns{secretservice.gov}) as well
as those charged with public health and safety (such as
\dns{fema.gov}, \dns{nih.gov}, and \dns{cdc.gov}).
At the state and local
level, virtually all primary state government domains (\eg, \dns{mass.gov})
  are vulnerable (including a broad range of congress, judiciary,
  and law enforcement domains in each state) and over 40\% of all \dns{.gov}
  domains used by cities.\footnote{We have not broadly examined domains representing government offices outside the US, but we note that both \dnsfn{gchq.gov.uk} and \dnsfn{ncsc.gov.uk} are also vulnerable.}

\paragraph{Example}
Figure~\ref{fig:open_forwarding_attack_mechanism} shows an example of
this attack using Outlook as the forwarding service.
An attacker starts by creating a personal account for
forwarding (\dns{attacker@outlook.com}), adding the spoofed address
(\dns{bush@state.gov}) to the account's ``allowlist'' (thereby
preventing any quarantining by Outlook), and configuring the account to
forward all email to the desired target (\dns{victim@gmail.com}). In this case, the spoofed domain \dns{state.gov} includes
Outlook's SPF record (\dns{spf.protection.outlook.com}) into its own SPF record and has a DMARC policy of \textsc{Reject}. Next, the attacker forges an email that purportedly originates from \dns{state.gov} and sends it to their personal Outlook account. Normally, Outlook would quarantine this email because it fails DMARC validation (\S~\ref{subsubsec:quarantine_instead_of_reject}). However, since the spoofed address is present in the account's allowlist, this configuration overwrites the quarantine decision (\S~\ref{subsubsec:whitelist}), and as a result, Outlook would forward the spoofed email to the target.

As per Outlook's MFEF forwarding implementation,
\textsc{MAIL FROM}
is rewritten to match the
\textsc{FROM} header, \dns{bush@state.gov} in our example. Finally, the recipient's mail server receives the forwarded email
and performs authentication checks.
From the recipient's perspective,
the spoofed email passes SPF validation because the
\textsc{MAIL FROM} domain (\dns{state.gov}) lists Outlook's SPF
information in its SPF record, and the forwarding configuration
arranged by the attacker ensures that the recipient receives this
spoofed email from Outlook's servers.
Moreover, this attack also
ensures that DMARC's alignment check succeeds because the
\textsc{MAIL FROM} and \textsc{FROM} domain are both \dns{state.gov}.
We validated this attack in practice, consistently sending
spoofed email messages such as the example
shown in Figure~\ref{fig:open_forwarding_attack_screenshot}
to our own Gmail account, where it was delivered to the inbox without warning.\footnote{
Note that we did discover some exceptions in our experiments.
For a small set of high-profile domains that have a DMARC policy of Reject
(\eg, \dnsfn{aa.com}, \dnsfn{foxnews.com} and \dnsfn{ikea.com}), Outlook would
quarantine spoofed email regardless of whether users have added the
spoofed address to their account's allowlist (Section~\ref{subsubsec:quarantine_instead_of_reject}).
We surmise that Outlook applies special protections for a set of high-profile or frequently spoofed domains.
} 
In addition to Outlook, this attack also succeeds with
iCloud, Freemail, Hushmail, Mail2World and Runbox.

\subsection{Abusing Relaxed Forwarding Validation}
\label{subsec:attack_relaxed_forwarding_validation}

The second attack exploits the fact that many email providers apply relaxed validation policies to forwarded mail (\S~\ref{subsubsec:relaxed_validation}), particularly when messages arrive from well-known mail providers.
When combined with open forwarding, an attacker can abuse this behavior
to spoof email from any domain that has a DMARC policy of
\textsc{Quarantine} (or \textsc{None}) to any mail server that applies these relaxed measures (\eg, Gmail and Outlook).  Recall that, in the absence of forwarding, attackers cannot spoof email from a domain with a DMARC policy of \textsc{Quarantine}.
Provider-specific defenses, such as when Outlook quarantines any email that fails DMARC (\S~\ref{subsubsec:dmarc_none}), will also stop such direct, single-hop attacks.

\begin{figure*}[t]
\centerline{\includegraphics[width=\textwidth]{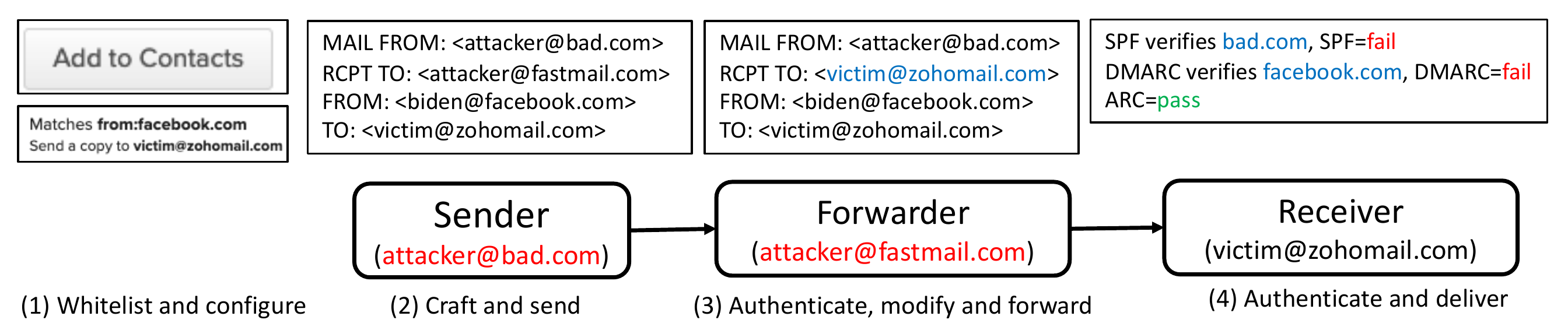}}
\centering
\caption{Example attack that exploits Zoho's
vulnerable ARC implementation and  open forwarding to
  spoof email from arbitrary domains to any Zoho recipient (\S~\ref{subsec:attack_zoho_arc}).}
\label{fig:mech_zoho_arc}
\end{figure*}

\paragraph{Scope}
As described earlier in Section~\ref{subsubsec:relaxed_validation}, Gmail and Outlook use relaxed validation checks for forwarded email.
We find that an adversary can mount this attack against users with
Gmail/Outlook email accounts as well as users who use GSuite and Outlook 365 for email services.\footnote{Mail.ru also uses relaxed validation, but since it is only applied to email forwarded via Gmail, which does not allow open forwarding, this attack does not work for Mail.ru.}

\paragraph{Example}
Figure~\ref{fig:mech_gmail_via_outlook} illustrates the steps of this
attack using an example where the adversary creates a personal Outlook
account to forward spoofed email messages to Gmail recipients.  First,
the adversary selects a spoofed email address from a domain with a
DMARC policy of \textsc{Quarantine} or \textsc{None} (we use \dns{alipay.com} in this example, a prominent Chinese payment company), adds the address
to their forwarding account's allowlist, and configures their
forwarding account to send email to the victim (recipient).  Like the
first attack, the attacker then sends a message from this spoofed address
to their forwarding account, which is then forwarded to the recipient.

When the final recipient's mail server receives
the email, the server will observe that the email comes from a
``well-known'' provider, apply its relaxed validation checks, and
successfully deliver the email to the recipient's inbox (even though
the spoofed email fails normal SPF and DMARC checks).\footnote{Additionally, we note that Gmail would usually display a UI warning for forwarded email messages. However, no UI warning is displayed for this email due to a bug detailed in Appendix~\ref{subsec:ui_bug}.}

\subsection{Targeting ARC Vulnerabilities}
\label{subsec:attack_zoho_arc}
The third attack allows an adversary to deliver spoofed email messages
from arbitrary domains to Zoho users.  This attack exploits Zoho's
vulnerable implementation of the experimental Authenticated Received
Chain (ARC) protocol~\cite{ARCSpeci1:online}, which was first
documented by Shen et al.~\cite{shen2020weak}. Due to this bug, Zoho
incorrectly reads ARC headers and will deliver arbitrary email
messages with ARC headers added by providers such as Gmail and
Fastmail to the recipient's inbox without any warning.  However, we
show that this issue is not limited to interactions between Gmail and
Zoho customers.  We demonstrate how further issues, including the fact that
Zoho trusts and (incorrectly) reads ARC headers added by Fastmail
(Appendix~\ref{sec:arc_adoption_and_trust}), open forwarding
(\S~\ref{subsubsec:open_forwarding}), and several forwarding
assumptions (\S~\ref{subsubsec:quarantine_instead_of_reject},
\S~\ref{subsubsec:whitelist}), can be combined with the underlying ARC
vulnerability to allow an adversary to deliver spoofed email messages
from arbitrary domains to arbitrary Zoho users.

This attack again highlights the fact that email security protocols
are distributed and independently-configured components, where
vulnerable decisions by one party incur harm to downstream recipients
but not necessarily to their own users.  Notably, the actions taken
by one provider (e.g., Fastmail) can unexpectedly undermine the
security of users on another platform (e.g., Zoho).

\paragraph{Scope}
Our experiments show that this attack can target arbitrary users of Zoho, which is estimated to have more than 10 million users~\cite{Celebrat69:online}.



\begin{figure*}[t]
\centering
\includegraphics[width=1.7\columnwidth]{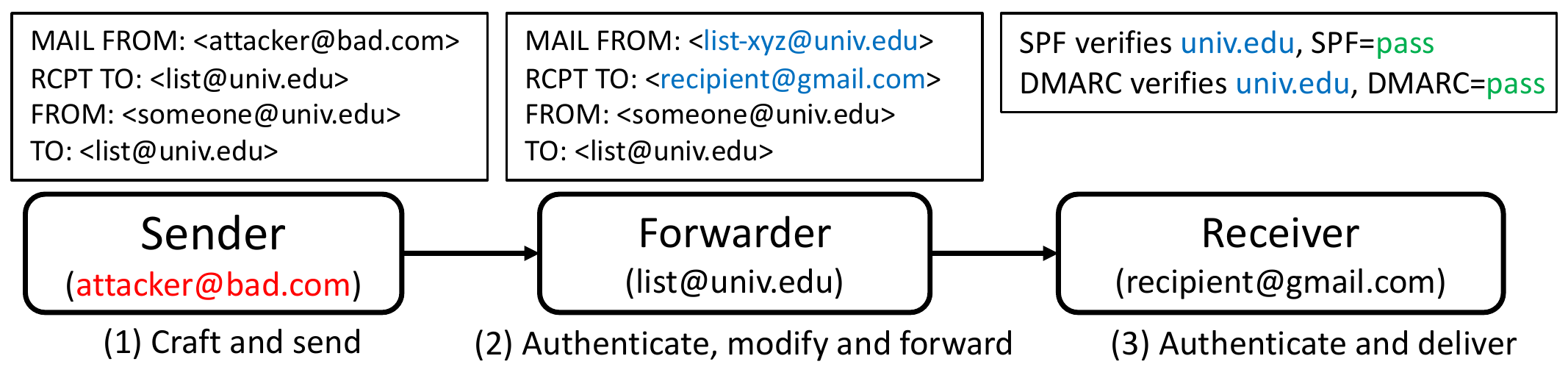}
\centering
\caption{Spoofed email attack that abuses mailing lists like Google Groups (\S~\ref{subsec:attack_none_mailing_list}).}
\label{fig:google_groups_mech}
\end{figure*}

\paragraph{Example}
Figure~\ref{fig:mech_zoho_arc} shows the mechanics of this attack
in the context of an attacker with a forwarding account on Fastmail who targets a recipient on Zoho.

First,
the adversary creates a Fastmail account for forwarding, adds their spoofed address
(\dns{biden@facebook.com}) to their allowlist, and configures their
account to forward all mail to the target user at Zoho
(\dns{victim@zohomail.com}).
Second, the adversary crafts and sends
spoofed email from their own servers (\eg, \dns{attacker@bad.com})
to their forwarding account at Fastmail.
Third, although this email will fail anti-spoofing validation, Fastmail will still faithfully forward it to the target user at Zoho due to the sender's presence on the user's allowlist (exploiting the security assumption discussed in \S~\ref{subsubsec:whitelist}). 
As part of the forwarding process,
Fastmail will modify the RCPT TO header and add corresponding ARC headers to the
spoofed email.
Finally, upon receiving the forwarded email, Zoho's mail server will perform DMARC validation.
Although the spoofed email will fail SPF and DMARC checks,
Zoho's vulnerable ARC implementation
will misinterpret the ARC headers that Fastmail attached (Appendix~\ref{sec:arc_adoption_and_trust}).
As a result, Zoho will treat the email as passing
DMARC and deliver the spoofed message to the victim's inbox. We end by noting that this attack would not have worked for domains with DMARC policy \textsc{Reject} had Fastmail rejected spoofed email messages addressed from such domains (\S~\ref{subsubsec:quarantine_instead_of_reject}).





\subsection{Abusing Mailing Lists}
\label{subsec:attack_none_mailing_list}

The final attack allows an adversary to abuse the forwarding process
used by mailing lists so that spoofed email, which would otherwise
fail DMARC authentication checks, successfully passes both SPF and
DMARC validation.
This attack targets domains with a weak DMARC policy of \textsc{None},
and exploits the way in which many mailing lists rewrite email headers during their forwarding process.
Concretely, this attack allows an adversary to abuse REM header rewriting (\S~\ref{sec:measure_forwarding_mechs_and_arc}) to launder spoofed email through mailing lists such that the forwarded email appears as if it originated from the legitimate sender, even though the original email fails DMARC authentication.\footnote{One such attack used to distribute phishing messages at our institution was part of the impetus for this study.}


\paragraph{Scope}
Our experiments show that attackers can
conduct this attack across all four popular mailing list services: Google
Groups, Mailman, Listserv, and Gaggle.

This attack only affects organizations that use a mailing list configured under their own domain name, and with a DMARC policy of \textsc{None} for their (sub)domain.
While these requirements appear restrictive, prior
work has found that many organizations
(such as major U.S.\ universities) have exactly this configuration~\cite{hutowardsunderstanding}.
Indeed, in querying \dns{.edu} and
\dns{.gov} domains, roughly 10\% of all \dns{.edu}
domains and 5\% of all \dns{.gov} domains are potentially susceptible to this attack.\footnote{As examples, Yale University operates \dnsfn{yale.edu} with a DMARC policy of \textsc{None} and hosts multiple mailing lists using Mailman, and the State of Washington operates a range of mailing lists using Listserv and whose \dnsfn{wa.gov} domain also has a DMARC policy of \textsc{None}.}

Additionally, for mailing lists like Gaggle that do not enforce DMARC
checks before forwarding (Appendix~\ref{subsec:no_dmarc}), this attack affects every organization using their services,
even when the domain has adopted stronger DMARC policies.

%

\paragraph{Example}
Figure~\ref{fig:google_groups_mech} describes an example of this attack
using Google Groups.
First, an attacker selects a target email address (\dns{someone@univ.edu}) to impersonate in their spoofed email,
and sends the spoofed message from their malicious server to the organization's mailing list (\dns{list@univ.edu}).
Although the email fails DMARC validation,
the mailing list will still accept the message (because \dns{univ.edu}
has a DMARC policy of \textsc{None}).
As part of REM forwarding, the mailing list will rewrite the
\textsc{MAIL FROM} header such that its domain matches the mailing
list's domain, and then forward the email to the list's members.
As a result, when a recipient's server receives the message it will
successfully pass SPF validation since the domain of the rewritten
\textsc{MAIL FROM} (\dns{univ.edu}) allows the mailing list to send on
its behalf.
Moreover, the spoofed message will also pass DMARC alignment checks,
since the rewriting performed during REM forwarding
ensures that the \textsc{MAIL FROM} and \textsc{FROM} domains are
identical.\footnote{Note that while our examples show this attack
using the organization's top-level domain, it is also effective for
any of the organization's subdomains (if the subdomains also have
DMARC policy \textsc{None}) due to DMARC's inherent relaxed alignment policy.}

During our experiments, we also observed that some mailing list
services, such as Gaggle, do not enforce DMARC policies at all
(Appendix~\ref{subsec:no_dmarc}).
This lack of enforcement allows the attack to succeed regardless of the spoofed domain's DMARC policy.
We provide more details in Appendix~\ref{sec:append_mailing_list_details}.

\section{Ethics and Disclosure}
\label{sec:disclosure}
When sending spoofed email messages in our experiments,
we took deliberate steps to avoid impacting any real users.
First, we only sent spoofed email messages to accounts that we created ourselves.
Second, we initially tested each attack by spoofing domains that we created and controlled for this research.
Once we established that our attacks could succeed using these test domains,
we ran a small set of experiments that spoofed email from real domains (to validate the absence of any unforeseen protection);
however, these email messages were only sent to our test accounts and did not spoof existing, legitimate email addresses from these domains.
Finally, all of our email messages contained innocuous text (e.g., ``a spoofed email'') that would not themselves cause harm.

We have disclosed all of the vulnerabilities and attacks to the
affected providers. As of the time of publication, we have received
affirmative feedback from all affected providers and we summarize our
current understanding of their present state here. Zoho has not only
patched the issue with their ARC implementation (also confirmed by
Wang et al.~\cite{wang2022revisiting}, who conducted their
measurements after the patch) and awarded us a bug bounty, but is also
further augmenting the security of its ARC implementation. Microsoft
confirmed the vulnerabilities (with severity ``Important'', the
highest severity assigned to email spoofing bugs) and awarded us a bug
bounty. They have partially fixed the issues by rejecting spoofed email
messages purporting to be from domains that have a DMARC policy of
\textsc{Reject}~\cite{hotmailreject}.
Gaggle confirmed the issues we flagged and stated that they would
start enforcing DMARC. Gmail fixed the issues we reported.  iCloud
partially fixed the issues we reported by not forwarding email
messages that fail DMARC authentication (except for domains with DMARC
policy \textsc{None}). Hushmail fixed the issues we reported by not
forwarding email messages that fail DMARC authentication. Freemail
fixed the issues we reported by not forwarding spoofed email messages
from domains that are their customers. Mail2World attempted to fix the
issues by using spam filters and remains vulnerable. Runbox did not
view the issues we reported as vulnerabilities. Instead, they consider
monitoring account activities post-complaints sufficient.


\section{Discussion and Mitigation}
We end by summarizing the root causes of the issues we discovered, and discuss potential mitigation strategies.

\subsection{Discussion}
In this work, we examine the complexities introduced by email forwarding to email security. We identify a diverse set of email forwarding mechanisms, assumptions, and features, and demonstrate how they can be combined together to perform evasion attacks. These attacks highlight four fundamental issues.

First, as already demonstrated in prior work and further highlighted in our paper, email security involves distributed, optional, and independently-configured components implemented by different parties. In such an architecture, the
``authenticity'' of an email is commonly determined by the party with the weakest security settings. While traditionally email is sent directly from sender to receiver, forwarding involves three parties instead of two and introduces an extra layer of complexity. As we have shown, a vulnerable forwarder can jeopardize the security of downstream recipients that do not have problematic configurations or implementations. This inversion of
incentives and capabilities naturally complicates
mitigating forwarding vulnerabilities.

A second problem is that email forwarding has never been fully standardized, despite the longevity and popularity of its use. A lack of standardization has led to ad-hoc implementation decisions, each making different assumptions.
This ad-hoc nature of implementations makes it challenging to perform both manual security analysis (analyzing individual implementation decisions is a non-trivial task even for experts) and automated testing (any such tool needs to account for the specific implementations of each provider).
While our large-scale empirical measurements have been able to reveal
the assumptions made by providers and their implications, it has
required substantial manual work.  This manual process is a reflection
of the fact that there exists no unified framework or standard for
implementing email forwarding.

A third issue is that email is a large, slowly-evolving ecosystem with a wide range of legacy systems and protocols that need to be accommodated.
One example we highlight is the ``outdated'' assumption made by SPF (\S~\ref{subsubsec:spf_incorporation}). When SPF was first designed in the early 2000s, it was common practice for each domain owner to maintain their own mail infrastructure. However, this assumption is obsolete in the modern era, as many domains outsource their email services to third-party providers such as Outlook and Google~\cite{liu2021s}. These large providers often share the same email infrastructure across all customers (both business and personal accounts), violating the assumptions made by SPF. 
To mitigate the risks this reality poses to SPF, providers 
usually prevent users from setting arbitrary values in their FROM header. However, past literature has shown that this defense is not always implemented correctly~\cite{chen2020composition}. We build on top of this prior work by identifying a new attack that can circumvent existing defenses through forwarding (\S~\ref{subsec:attack_open_forwarding}).

Last but not least, the intrinsic nature of email forwarding is to transparently send an existing message to a new address ``on behalf'' of its original recipient --- a goal very much at odds with the anti-spoofing function of protocols such as SPF and DMARC. As such, a range of ad-hoc decisions have been made to increase the deliverability of forwarded email messages, such as using the REM+MOD forwarding mechanism (\S~\ref{sec:measure_forwarding_mechs_and_arc}), treating forwarded messages specially (\S~\ref{subsubsec:relaxed_validation}), and adding DKIM signatures to forwarded messages (\S~\ref{subsubsec:unsolicited_dkim}). As we have demonstrated, 
these decisions can fail to foresee unexpected interactions that lead to vulnerabilities, even with a lot of deliberation.

\subsection{Mitigation}
The attacks we demonstrate highlight the complicated interactions
between email forwarding and existing anti-spoofing mechanisms. We start by reviewing short-term mitigations that could reduce some of the most significant risks we have uncovered. We then discuss challenges in developing more comprehensive solutions, which would require significant changes in either
protocol or operational practices.

A core issue we highlight in this paper is the ability to forward spoofed email messages to arbitrary recipients, a critical element in each
of the first three attacks in Section~\ref{sec:attacks}. To mitigate this issue, providers could either block spoofed email messages from being forwarded, or enforce that a forwarder can only forward to accounts under their control by requiring explicit confirmation (similar to
the online domain validation used by modern certificate authorities). However, we note that either approach comes with a usability tradeoff, and different providers make choices based on their considerations. Indeed, providers like Gmail and Mail.ru opted for the former option, while others like iCloud and Hushmail opted for the latter.




As well, we advocate that providers
should enforce a domain's DMARC \textsc{Reject} policy when specified, rather
than substituting a weaker policy.  If Outlook rejected spoofed
email messages from such domains, the impact of the first attack
exploiting SPF incorporation
would
narrow substantially. We understand that Outlook has plans to take such action in the future~\cite{hotmailreject}.

Unfortunately, all the defenses described above reflect a
case of misaligned
incentives: the recipients of spoofed email (\eg, spam and phishing)
cannot implement this change, but instead need to rely on the entire
ecosystem of providers and forwarding services to adopt such defenses. 

Email providers can also mitigate the second attack (\S~\ref{subsec:attack_relaxed_forwarding_validation}) by eliminating
relaxed validation policies.  This approach would protect their users
from receiving spoofed email without relying on changes by other
platforms or services.  However, to prevent benign forwarding from
breaking will likely require providers to then implement ARC
validation (which in turn places ARC implementation requirements on
external forwarders).


For the final attack (\S~\ref{subsec:attack_none_mailing_list}) that exploits mailing lists, potential mitigations trade usability for security.
First, list owners can turn on message moderation and set their mailing lists to be private.
While these measures increase the difficulty of performing email spoofing attacks, they do not rule out the attack entirely. A dedicated attacker might
nonetheless identify a member of the mailing list and craft an email
that fools a list's moderator.
Second, some mailing list services, such as Listserv, support confirm-before-send~\cite{OnmyLIST7:online}, which requests confirmation from the (true) sender address before delivery.  While this mechanism would impose significant overheads in general, these costs might be acceptable by limiting this confirmation requirement to incoming email that fails DMARC authentication checks.

In addition to the short-term mitigations mentioned above that are
specific to forwarding, others~\cite{chen2020composition,shen2020weak}
have proposed solutions such as improving UI notification, building
better testing tools, and revising RFC standards, which are also
important to consider. Additionally, the newly proposed ARC protocol
may also help mitigate some of the issues we have uncovered. However,
ARC is still in the early stages of development and deployment, 
its details are yet to be fleshed out and its effectiveness in
practice remains to be seen.

Lastly, we note that comprehensively fixing email forwarding would
require a more fundamental set of changes (e.g., redesigning the
entire suite of email security protocols), which will face significant
deployment challenges given the current state of the email ecosystem.
Chief among these challenges is that any new solution designed to fix
forwarding must address backwards compatibility, a task complicated by
email's forty-year-old ecosystem of varied protocols, implementations
and use cases.  Specifically, one must carefully consider how any new
approach interacts and interoperates with existing systems (e.g., mail
providers and filtering service providers) and protocols (e.g., SPF,
DKIM and DMARC).  While security might be enhanced by embracing a
single standard approach to forwarding (e.g., when a message
should be forwarded, what forwarding mechanisms should be used, what
information should be added to forwarded messages, and how the
receiving account should be verified), any such choice will inevitably
  align well with certain providers and conflict with those whose
  existing services have made different choices or who operate under
  different threat models.  Finally, it is not enough to merely
  standardize new protocols, but one must then also incentivize and
  coordinate their universal deployment and operation.  Thus, while
  such an aspirational goal is worthy of attention, it seems likely
  that email will continue to benefit from incremental and reactive
  improvements, such as those discussed earlier, for some time yet.

\section{Conclusion}
\label{sec:discussion}
Internet-based email has been in use since the early 1970s
and the SMTP protocol has been in use since 1980.  It is arguably the
longest-lived text-based communication system in wide use.
Unsurprisingly, its design did not anticipate the range of challenges
we face today and, because of its central role, we have been forced
to upgrade email protocols slowly and with deference to a wide range
of legacy systems and expectations. Perhaps nowhere is this more clear
than around the issue of authentication.  Email protocols have no
widely-used mechanism for establishing the authenticity of sender
addresses, and thus we have focused on authenticating the domain
portion of the email address (largely motivated by spam and phishing).

In this work, using large-scale empirical measurements of 20 prominent email forwarding services, we identify a diverse set of email forwarding mechanisms, assumptions, and features, and demonstrate how they can be combined together to perform four types of evasion attacks. While we are the first academic paper to document these attacks, retrospectively examining Mailop~\cite{Mailop96:online}, a prominent mailing list for mail operators, we have also found traces~\cite{RealTraces} of real-world attacks that are similar to what we reported in this paper. 

The attacks we document exploit four kinds of problems. One fundamental issue is that email security protocols are
distributed, optional, and independently-configured components. This creates a large and complex attack surface with many
possible interactions that cannot be easily anticipated or
administered by any single party. A second problem is that email forwarding was never standardized, leading to ad-hoc implementation decisions that might be vulnerable. A third problem is that protocol assumptions for SPF are grounded at a
point in time and have not been updated as practices have changed. Domains now out-source their mail service to large providers that share mail infrastructure across customers, undermining assumptions made in the design of SPF. Lastly, the intrinsic nature of email forwarding is to transparently send an existing message to a new address ``on behalf'' of its original recipient. 
This creates complex chain-of-trust issues that are at odds with implicit assumptions that mail is sent directly from sender to receiver. Indeed, it is this complication that has driven the creation of ARC.

While there are certain short-term mitigations (\eg, eliminating the use of
open forwarding) that will significantly reduce the exposure to the
attacks we have described here, ultimately email requires a more
solid security footing if it is to effectively resist spoofing
attacks going forwards.

\section*{Acknowledgments}
We thank our anonymous reviewers for their insightful and constructive suggestions and feedback.
We thank Nishant Bhaskar, Cindy Moore, and Jennifer Folkestad for their operational support. We thank Stewart Grant for proofreading the paper. We thank Brad Chen for collecting feedback from Google. We thank John Levine and Weihaw Chuang for their comments on the paper.  Funding for this work was provided in part by National Science
Foundation grants CNS-1705050 and CNS-2152644, the UCSD CSE
Postdoctoral Fellows program, the Irwin Mark and Joan Klein Jacobs
Chair in Information and Computer Science, and operational support
from the UCSD Center for Networked Systems as well as the Twente University Centre for Cybersecurity Research (TUCCR).



\bibliographystyle{IEEEtran}
\bibliography{ref}


\appendices

\newpage
\section{ARC Adoption in the Wild}
\label{sec:arc_adoption_and_trust}
\begin{table}[t]
  \centering
\begin{tabular}{l|llll}
  \toprule
  & \multicolumn{3}{c}{\textbf{Added by}} \\
\textbf{Received by} & Gmail & Zoho & Fastmail & Pobox\\
\midrule
Gmail    & \checkmark     &                &       \checkmark  & \checkmark\\
Outlook  & \checkmark     &               &         & \\
Zoho     & \checkmark     &     & \checkmark      & \checkmark    \\
Fastmail & \checkmark     & \checkmark    & \checkmark   & \checkmark\\
Pobox    & \checkmark     &               & \checkmark   & \checkmark\\
\bottomrule
\end{tabular}
\caption{Trust of ARC headers between providers.\label{tab:trust_of_arc_between_providers}}
\end{table}

\label{sec:appendix_arc_measurement}
Five email providers (Gmail, Outlook, Zoho, Fastmail, and Pobox) and two mailing list services (Google Groups and Mailman) implement ARC validation.
Because ARC is still an experimental protocol, many email providers only evaluate ARC headers added by a small set of other providers whom they trust~\cite{Senderau57:online}.
Based on measurements through test accounts that we created, Table~\ref{tab:trust_of_arc_between_providers} shows the ARC trust relationships among the providers we tested:
Gmail trusts ARC headers added by Fastmail, Pobox and itself; Outlook trusts ARC headers added by Gmail; Zoho trusts ARC headers added by Gmail, Fastmail, Pobox and itself;  Fastmail trusts ARC headers added by Gmail, Zoho, Pobox and itself; and Pobox trusts ARC headers added by Gmail, Fastmail and itself.

We also note two details.
First, we cannot test whether other providers trust ARC headers added by Outlook. ARC headers are only evaluated when a forwarded email fails DMARC authentication checks.
However, in our experiments, Outlook only adds ARC headers to certain email messages that will pass DMARC authentication checks after forwarding.
This design prevents us from testing which providers trust Outlook's ARC headers.
Second, our results suggest that Zoho does not add ARC headers when forwarding messages internally between Zoho accounts, so we leave that cell empty in the table.

\section{Additional Implementation Errors}
\label{sec:implementation_errors_additional}
We detail two other implementation errors identified during our measurement. These two errors play minor roles in the attacks we discover.
\subsection{Ignoring Security Protocols}
\label{subsec:no_dmarc}
Systems that assume every email is legitimate tend not to enforce DMARC. Instead, they will permissively accept
email messages, even if they fail DMARC authentication checks and the
purported FROM domains have a stricter DMARC policy of \textsc{Quarantine} or
\textsc{Reject}. Such assumptions can create serious security issues.

From our experiments, we found
one mailing list provider (Gaggle) and three mail providers (Freemail, Mail2World and Runbox) that do not enforce
DMARC.
Beyond delivering spoofed email to users and allowing spoofed email to be forwarded, we
show later (Section~\ref{subsec:attack_none_mailing_list}) that
incorrect enforcement, combined with the standard forwarding
modifications that Gaggle applies to email headers, allows an attacker
to abuse Gaggle. Spoofed email messages that initially fail
DMARC authentication will receive a fresh set of headers that
correctly pass SPF and DMARC validation after they are forwarded
through a Gaggle-operated mailing list.

\begin{figure}[t]
  \centering
{
    \setlength{\fboxsep}{0pt}
    \setlength{\fboxrule}{0.5pt}
    \fbox{\includegraphics[width=\columnwidth]{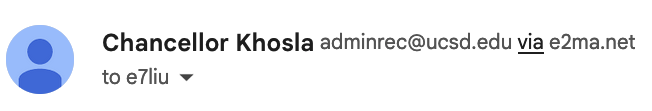}}
}
  \caption{Gmail annotating the sending address of an email. 
}
\label{fig:gmail_via}
\end{figure}

\subsection{Gmail UI Bug}
\label{subsec:ui_bug}
After a receiver accepts and processes email,
the user's mail user agent (MUA) displays the message for viewing.
Thus, MUAs and their UI warnings serve as the last line of defense against spoofed email messages.
However, previous work~\cite{hu_end--end_nodate,shen2020weak,chen2020composition} has found multiple security issues in various MUAs, especially on mobile platforms.
In our experiments, we focus on the native MUAs (web interfaces) provided by the nine email platforms in our study.
These MUAs not only have widespread usage, as the default MUA for many users,
but are also maintained by the email providers and tend to have better security practices.

Among all native MUAs, only Gmail, Onet and Zoho have implemented warning
systems that display UI indicators when an email is forwarded or fails
DMARC authentication.  Gmail, for instance, annotates the
sending address (e.g., \dns{adminrec@univ.edu} \textit{via} \dns{e2ma.net} as shown in Figure~\ref{fig:gmail_via}).


However, we observed a bug in Gmail's warning system for a subset of forwarded
messages. In particular, Gmail does not display an indicator for a forwarded
email message if (1) it does not contain any DKIM headers, and (2) it has the
same domain in both the \textsc{MAIL FROM} and \textsc{FROM} headers.
This policy does not pose a problem in single-sender email settings, because adversaries still need to bypass SPF and DMARC.
However, we present a new attack that uses this bug in conjunction with forwarding and other vulnerabilities to deliver spoofed email messages that look no
different than legitimate messages (Section~\ref{subsec:attack_relaxed_forwarding_validation}).

\section{Additional Attack Screenshots}
We ran a small set of experiments that spoofed email impersonating real domains to validate the attacks described in Section~\ref{sec:attacks}.
Our experiments confirmed that these attacks succeed.
Below, we present the screenshots of spoofed email messages successfully delivered to users' inboxes.

\begin{figure}[t]
  \centerline{\includegraphics[width=\columnwidth]{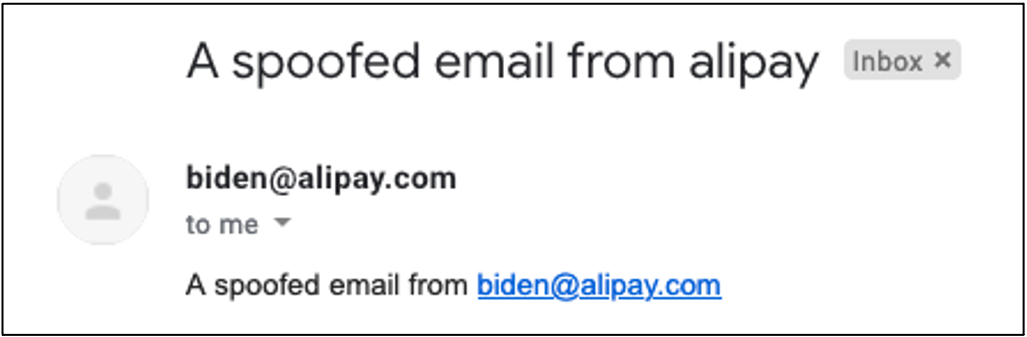}}
  \centering
  \caption{Email spoofing \dns{biden@alipay.com} via Outlook.}
  \label{fig:ss_gmail_via_outlook}
  \end{figure}

For the attack described in Section~\ref{subsec:attack_relaxed_forwarding_validation}, Figure~\ref{fig:ss_gmail_via_outlook} shows a spoofed message forwarded via a personal Outlook account to a Gmail account we created, and delivered to the recipient's inbox without any security warnings.
The spoofed address in this example impersonates a sender at \dns{alipay.com} (a prominent Chinese payment company with a DMARC policy of \textsc{Quarantine}).

\begin{figure}[t]
  \centerline{\includegraphics[width=\columnwidth]{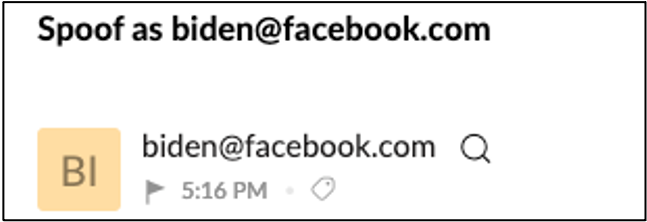}}
  \centering
  \caption{Email spoofing \dns{biden@facebook.com} via Fastmail.}
  \label{fig:ss_zoho_arc}
  \end{figure}

For the attack described in Section~\ref{subsec:attack_zoho_arc}, Figure~\ref{fig:ss_zoho_arc} illustrates that this attack succeeds
without any security warnings, even though the spoofed domain in our
experiment, \dns{facebook.com}, has a DMARC policy of \textsc{Reject}.

\section{Additional Details for the Attack in Section~\ref{subsec:attack_relaxed_forwarding_validation}}
\label{sec:append_change_behavior_details}
This section makes three additional observations about the attacks on
providers with relaxed forwarding validation as described in
Section~\ref{subsec:attack_relaxed_forwarding_validation}.

First, in addition to forwarding from personal Outlook accounts, an
adversary can also forward from personal accounts with other providers
(\eg, Fastmail) to Gmail recipients.
As mentioned earlier in Section~\ref{subsubsec:relaxed_validation}, there is an additional caveat: the \textsc{TO} header of the spoofed email cannot be the same as victim's email address.
An astute recipient might see that the \textsc{TO} field corresponds to someone else's email account, and become suspicious about the email's validity.
To reduce suspicion in this case, the adversary can set the
human-readable name portion of the email's \textsc{TO} header to
``me'' or the victim's name (while keeping the address different).

Second, adversaries need to leverage popular mail providers as forwarders in this attack; they cannot exploit relaxed forwarding validation by using their own servers as forwarders.
In our experiments, we tested using both a personal Outlook account as well as a mail server we controlled as forwarders.
The first version results in successful attacks delivered to user inboxes, but the latter did not.
We suspect this outcome is because our mail server domain has lower reputation than Outlook's mail servers.

\begin{figure}[t]
\centerline{\includegraphics[width=\columnwidth]{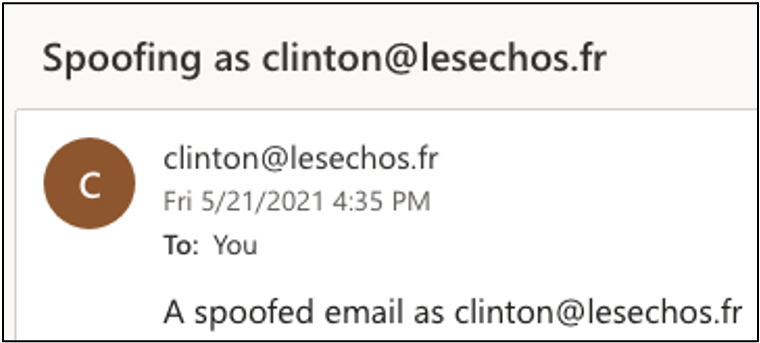}}
\centering
\caption{Spoofed email message taking advantage of Outlook's relaxed forwarding validation policy.}
\label{fig:attack_none_ms}
\end{figure}

Finally, in Section~\ref{subsec:attack_relaxed_forwarding_validation}, we demonstrate the attack against a recipient using Gmail. A similar attack can be mounted against Outlook recipients by
forwarding via a personal Fastmail account. This attack allows an
adversary to spoof email messages from many domains that have a DMARC
policy \textsc{None} to arbitrary Outlook recipients.\footnote{Similar to the caveats described in
Section~\ref{subsubsec:quarantine_instead_of_reject}, we observe that Outlook applies
additional restrictions to a small set of high-profile domains with a
DMARC policy of \textsc{None} (e.g., \dns{citizensbank.com}), which blocks the delivery of spoofed emails from these domains.}
Figure~\ref{fig:attack_none_ms} shows a spoofed email message
forwarded via a personal account to an Outlook account we created, and
delivered without any security warnings. The spoofed FROM header in
this example impersonates a sender at \dns{lesechos.fr} (a French
financial newspaper with a DMARC policy of \textsc{None}).

\section{Additional Details for the Attack in Section~\ref{subsec:attack_zoho_arc}}
\label{sec:appendix_zoho_attack_details}
Adversaries can broaden the scope of the attack described in Section~\ref{subsec:attack_zoho_arc} by using a forwarding
account at any email provider that Zoho trusts for ARC purposes,
including Gmail, and routing their spoofed email through multiple forwarding hops.
In particular, an attacker can obtain ARC headers in one forwarding hop via Gmail, and then bypass Gmail's lack of open forwarding by forwarding the email through a second account that does allow open forwarding (\eg, Outlook).
For example, first, the attacker would send their spoofed email message
to their Gmail account, which they configured to forward to their
malicious Outlook account.
During the forwarding process Gmail will
attach a set of ARC headers to the email message.
Next, the spoofed email will arrive at their malicious Outlook account, which then forwards the email to any arbitrary Zoho recipient (because Outlook
supports open forwarding).
This forwarded email message will contain Gmail's attached ARC headers, enabling the attack to successfully pass DMARC validation checks as a result of Zoho's vulnerable ARC implementation.

Using our test accounts, we validated that this multi-hop attack variation
successfully delivers spoofed messages to the inbox of a Zoho recipient
without any warnings.

\section{Additional Details for the Attack in Section~\ref{subsec:attack_none_mailing_list}}
\label{sec:append_mailing_list_details}


\begin{figure}[t]
\centering
\includegraphics[width=\columnwidth]{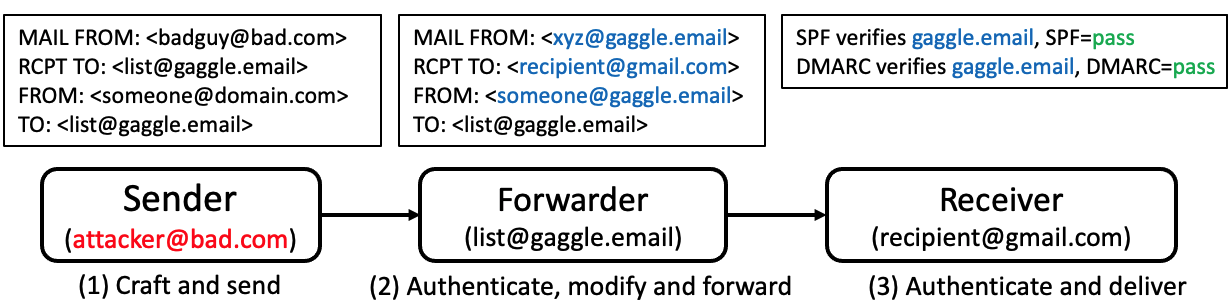}
\centering
\caption{Attack flow for Gaggle.}
\label{fig:gaggle_email_mech}
\end{figure}

In addition to the attacks described in
Section~\ref{subsec:attack_none_mailing_list}, we found
additional attack variants related to Gaggle.
Figure~\ref{fig:gaggle_email_mech} shows an example of an attack that
abuses Gaggle's use of REM + MOD forwarding
(Section~\ref{sec:measure_forwarding_mechs_and_arc}). This attack works
regardless of the DMARC policy of the spoofed address's domain.  First, an
attacker chooses an address to spoof (\dns{someone@foo.com}) that is
allowed to send to a mailing list on a vulnerable provider
(\dns{list@gaggle.email}), and sends a spoofed email message
purporting to come from that address.  This spoofed email will fail
DMARC validation, but because Gaggle does not enforce DMARC
(Appendix~\ref{subsec:no_dmarc}), it will forward the email to the
mailing list's recipients as normal (Stage 2).  Since Gaggle uses a
REM + MOD forwarding process, it will rewrite the \textsc{MAIL FROM}
header to use the mailing list's domain (\eg, a
new \textsc{MAIL FROM} address of \dns{xyz@gaggle.email}).
Finally, when the spoofed email message arrives at the recipient's mail server,
it will properly pass SPF validation and DMARC alignment checks:
the rewritten \textsc{MAIL FROM} domain allows the mailing list to send on its behalf, and the domain matches the \textsc{FROM} address's domain (\dns{gaggle.email}).


Additionally, we note that mailing list software such as Listserv and
Mailman require a backend MTA.  In our experiments we used Postfix
with DMARC turned on, a configuration which follows good security
practice.  However, in practice many organizations might not use this
configuration because many MTAs (including Postfix) do not enforce
DMARC by default.  In these cases, the attacker can spoof email from any
target domain, regardless of its DMARC policy, much like the attack
against Gaggle.


\end{document}